\documentclass[twocolumn]{article}

\usepackage{hyperref}
\usepackage{placeins}
\usepackage{adjustbox}
\usepackage{sectsty}
\usepackage{helvet}
\usepackage{appendix}
\usepackage{threeparttable}

\usepackage{tikz}
\usepackage{array} 
\usetikzlibrary{positioning,arrows.meta}

\usepackage[%twoside,%
letterpaper,
			,includeheadfoot,%
			layoutsize={8.125in,10.875in},%
                layouthoffset=0.1875in,%
                layoutvoffset=0.0625in,%
                left=38.5pt,%
                right=43pt,%
                top=43pt,% 10pt provided by headsep
                bottom=44pt,%
                headheight=0pt,% No Header
                headsep=-2pt,%
                columnsep=0.3125in,
                footskip=25pt,
                marginparwidth=38pt]{geometry}
\usepackage{authblk}
\usepackage{booktabs}
\usepackage{array}
\usepackage{graphicx}
\usepackage{multirow}
\usepackage{amsmath}

\usepackage{amssymb} 
\usepackage{float}
\usepackage{subfig}
\usepackage{pdfpages}
\usepackage{adjustbox}
\usepackage{blindtext}
\usepackage[numbers,sort&compress,merge,square]{natbib}
\usepackage{fancyhdr}
\fancypagestyle{firstpage}{%
  \fancyhf{}%
  \fancyfoot[C]{\footnotesize
    This version has not been peer reviewed. Comments and feedback are welcome.}%
}

\usepackage{caption}
\title{\LARGE Identifying Latent Intentions via Inverse Reinforcement Learning in Repeated Linear Public Good Games}

\date{}

\author[1]{Carina I. Hausladen}
\author[2]{Christoph Engel}
\author[2]{Marcel Schubert}

{
\footnotesize
\affil[1]{ETH Zurich, Computational Social Science}
\affil[2]{Max Planck Institute for Research on Collective Goods}
}

\begin{document}

\twocolumn[
  \begin{@twocolumnfalse}
    
\maketitle
\thispagestyle{firstpage}

\begin{abstract}
\noindent
Behavior in repeated public goods games continues to challenge standard theory: heterogeneous social preferences can explain first-round contributions, but not the substantial volatility observed across repeated interactions. Using 50{,}390 decisions from 2{,}938 participants, we introduce two methodological advances to address this gap. First, we cluster behavioral trajectories by their temporal shape using Dynamic Time Warping, yielding distinct and theoretically interpretable behavioral types. Second, we apply a hierarchical inverse Q-learning framework that models decisions as discrete switches between latent cooperative and defective intentions.
This approach reveals a large ($21.4\%$) and previously unmodeled behavioral type—Switchers—who frequently reverse intentions rather than commit to stable strategies. At the same time, the framework recovers canonical strategic behaviors such as persistent cooperation and free-riding.
Substantively, recognizing intentional volatility helps sustain cooperation: brief defections by Switchers often reverse, so strategic patience can prevent unnecessary breakdowns.
\vspace{0.50cm}
\end{abstract}
  \end{@twocolumnfalse}
]

\renewcommand{\thefootnote}{\fnsymbol{footnote}}
%\footnotetext[0]{Code not yet publicly available}
\footnotetext[0]{Corresponding author: carinah@ethz.ch}
\renewcommand{\thefootnote}{\arabic{footnote}}
\setcounter{footnote}{0}
%%%

\section{Introduction}

Cartels owe their fragility to a simple economic fact: cooperation is difficult to sustain when individual incentives point the other way. Economists often model this tension as a repeated public-goods dilemma. 
This framework is particularly apt for tacit collusion,  where firms coordinate without explicit agreements. Here, defection remains individually tempting, but implicit coordination mechanisms—such as delayed punishments or ``noisy" attempts to restore discipline can sustain cooperative outcomes \citep{engel2015tacit}.
Similar dynamics arise well beyond cartels. Many financial markets are characterised by sequences of interdependent actions:
Market-making, for instance, depends on a continuous stream of liquidity provision that is costly to supply yet easily exploited \citep{hendershott2011does}. 
Banks engage in funding-liquidity management, watch one another’s positions and adjust their own in response \citep{drehmann2013funding}. 
Clearing members in derivatives markets face shared margin obligations, where one firm’s stress can force others to contribute more \citep{king2022}. 
These environments, such as public-goods games, produce time series of contributions and withdrawals that are mutually conditioned.
The laboratory model thus provides a useful lens for studying their behaviour.

In this context, standard theory predicts the tragedy of the commons. Everybody maximizes individual profit and exploits the socially-minded choices of others. If community members interact repeatedly, but it is known when interaction will stop, the gloomy prediction still holds \citep{Kreps1982}. A robust experimental literature shows that, in the aggregate, results look different. In a standard symmetric linear public good game, average contributions typically start considerably above zero, but tend to decline over time \citep{Ledyard1995,Zelmer2003,Chaudhuri2011,burlando2005heterogeneous, Camerer2006, Fehr2002, fischbacher2010social}. A substantial theoretical literature rationalizes this result by introducing some form of social preferences into the utility function \citep{Camerer2006, Fehr2002, fischbacher2010social, Norenzayan2008}, whereas other scholars attribute cooperation to confusion, arguing that participants misinterpret the game \citep{Andreoni1995, Houser2002, BurtonChellew2016}. Efforts to disentangle cooperation driven by social preferences from that motivated by confusion have not reached a consensus. A widely accepted view posits that observed contributions stem from a combination of confusion and social preferences \citep{BurtonChellew2013, Bayer2013, Ferraro2010}. Some work argues that social preferences do not account for human cooperation at all, but instead, confusion does \citep{BurtonChellew2016}. Very recently, this conclusion was challenged \citep{wang2024confusion}.

Thus, the question of whether and how social preferences explain cooperative behavior in social dilemma games remains unresolved. This, in particular, holds for the development of contributions over time. 
Previous research has made important progress by analyzing experimentally generated gameplay data through partitioning approaches that interpret resulting clusters through the lens of established behavioral types. These methods have yielded valuable insights into heterogeneity in social preferences. However, two methodological considerations suggest opportunities for refinement.

First, partitioning typically relies on {\em pointwise alignment}. While this approach may be suitable for macroeconomic settings where policy changes affect all firms simultaneously, gameplay data tends to exhibit behavioral reactions that are affected by the idiosyncratic experiences in the randomly composed group. We propose using dynamic time warping, a method specifically designed to handle {\em shifted time series}, for partitioning gameplay data.

Second, the interpretation of partitions has been constrained by strong theoretical priors taken from preference types identified in static settings (e.g., modeling conditional cooperators as participants who match others' cooperation levels). Essentially, this literature {\em presupposes knowledge of players' reward functions}. Given the persistent presence of unexplainable noise or ``other" clusters, we instead argue for explicitly modeling {\em reward functions as an estimation problem}. 
%We propose employing Inverse Reinforcement Learning (IRL), where reward functions are not predefined but treated as part of the fitting objective.

\section{Literature}
\label{sec:literature}

\paragraph{Behavioral Types}
Categorizing human behavior in social dilemma games has long been of interest to the social sciences. 
For example, \citet{fischbacher2001conditionally} introduced the strategy method as a tool to classify individuals' predispositions toward cooperation in linear public good games.\footnote{The strategy method has also been criticized for risking confusing social preferences with conditional cooperation \citep{burton2022strategy}.}  
This method isolates preferences from the potential influence of learning dynamics or other mechanisms that may emerge during gameplay. 
Building on this foundational work, several subsequent studies \citep{abeler2015self, aimone2013endogenous, cubitt2017conditional, dariel2014cooperators, fischbacher2010social, fischbacher2001conditionally, fischbacher2012behavioral, fischbacher2014heterogeneous, fosgaard2014understanding, gaechter2017reciprocity, herrmann2009measuring, kamei2012locality, kocher2008conditional, makowsky2014playing, thoni2012microfoundations, van2014implementing, volk2012temporal, weber2018dispositional,Fallucchi2019}
have used the strategy method to identify distinct behavioral types in the very same game. A meta-analysis of these studies \citep{thoni2018conditional} found that the latter consistently report the existence of three types: free riders (who contribute nothing, 19.2\%), conditional cooperators (whose contributions depend on others' behavior, 61.3 \%), and triangle cooperators (10.4\%).

\begin{table*}
\centering
\caption{Behavioral Type Classifications Across Studies}
\label{tab:behavioral_types}
\resizebox{\textwidth}{!}{% 
\begin{threeparttable}
\begin{tabular}{lcccccc}
\toprule
 & \multicolumn{3}{c}{Conditional/Reciprocal Types} & \multicolumn{2}{c}{Selfish/Strategic Types} & {\bf Other} \\
\cmidrule(lr){2-4}\cmidrule(lr){5-6}\cmidrule(lr){7-7}
Study & Near-rational & Strong CC & Weak CC & Own-Max./Gamesmen & Fatalist & {\bf Confused/Others} \\
\midrule
Houser (2004) & 56\% & -- & -- & -- & 20\% & 24\% \\
Bardsley (2007) & -- & -- & -- & Free-riders: 25\%\tnote{a} & Altruists: 6\%\tnote{a} & Trembles: 18.5\%\tnote{b} \\
Fallucci (2021) FM\tnote{d} & 58\% (Recip.) & -- & -- & 9\% (Gamesmen) & -- & 32\% \\
Fallucci (2021) RM\tnote{d} & 25\% (Recip.) & -- & -- & 52\% (Gamesmen) & -- & 22\% \\
Fallucci (2019) & -- & 38.8\% & 18.9\% & 25.8\% (Own-Max.) & -- & 21.2\%\tnote{c} \\
\bottomrule
  \end{tabular}% <------ Don't forget this %
\begin{tablenotes}[flushleft]\footnotesize
\item \textbf{Notes}: All numbers and descriptions are reported directly from the respective cited studies. \textit{Near-rational}: Behave optimally in line with expected utility. \textit{Fatalist}: Fail to update beliefs properly, acting as if actions have no impact. \textit{Confused/Others}: High randomness or residual unclassified types. \textit{Strong Conditional Cooperators (CC)}: Match others' contributions approximately one-for-one. \textit{Weak CC}: Increase contributions less than proportionally to others' contributions. \textit{Own-Maximizers/Gamesmen}: Purely self-interested strategic players, contributing little or nothing unless beneficial. \textit{Reciprocators}: Respond positively to cooperative behavior.\newline
\item[a] Percentages reflect corrected estimations after accounting for trembles.
\item[b] Initial tremble probability, reflecting random decision errors, decreases significantly with experience.
\item[c] Combines unconditional high contributors (4.7\%) and residual (16.5\%).
\item[d] Fixed matching (FM); Random Matching (RM).
\end{tablenotes}
\end{threeparttable}
}
\end{table*}

Other research has also explored the classification of behavior in public good games, but with a focus on repeated interactions in {\em gameplay} data rather than data obtained through the strategy method \citep{burlando2005heterogeneous, kurzban2001individual, kurzban2005experiments, bardsley2007experimetrics, fischbacher2010social}. 
\citet{muller2008strategic} compare both approaches and find that gameplay data often aligns with the classifications derived from the strategy method.

Unlike the strategy method, which assumes a fixed decision rule, gameplay data requires modeling evolving strategies and decision noise. 
Proposed methods to form taxonomies in {\em gameplay data} from PGGs include Bayesian models \citep{Houser2004}, finite mixture models \citep{bardsley2007experimetrics}, clustering techniques \citep{bolle2021behavioral, Fallucchi2019}, Classifier-Lasso (C-Lasso) \citep{su2016identifying, BordtFarbmacher2019}, and Classification and Regression Trees \citep{Engel2020cart}.

% I. The other cluster
A critical point of comparison of these different approaches lies in how these papers accommodate behaviors that elude standard theoretical explanations.
Residual groups are termed ``confused" \citep{Houser2004}, ``various" \citep{Fallucchi2019}, ``others" \citep{fallucchi2021identifying}, or alternatively are captured through random tremble terms \citep{bardsley2007experimetrics}. 
Their size ranges from 18.5 to 32 percent (see \autoref{tab:behavioral_types}). 

These labels—confused, various, others—all implicitly frame the residual group as behavior that cannot (yet) be accommodated by existing theories. When the unexplained share is small, it is pragmatically reasonable not to introduce an additional type: with very few observations, there is simply not enough structure to interpret or model it. However, if nearly one-third of participants behave in ways that existing models fail to capture, it seems no longer appropriate to treat this as mere ``noise." Instead, it becomes natural to ask whether these behaviors reflect coherent, yet unmodeled, patterns that deserve explicit interpretation.

%Some degree of random behavior is undoubtedly present, but a share of up to 32 percent is striking and merits closer examination of whether it reflects genuine randomness or an artifact of the modeling approach.

% II handling temporal misalignment
A key observation shared across all those studies that propose methods to form taxonomies is the identification of a ``downward trend" in public good contributions.
This terminology inherently frames the data as time-series and, importantly, emphasizes trends over specific temporal locations of features such as peaks, which may vary due to idiosyncratic group dynamics. 
However, none of the above-cited papers explicitly addresses this temporal misalignment via their identification strategy.

\paragraph{Theory Driven to Data Driven Identification}
The above-cited methods can be placed along a spectrum from theory- to data-driven:

At the theory-driven end, finite mixture models are widely used \citep[e.g.,][]{bardsley2007experimetrics}, where specific behavioral strategies such as free-riding or conditional cooperation are explicitly modeled. 
Bayesian mixture models \citep[e.g.,][]{Houser2004} offer greater flexibility by estimating individual decision parameters through Bayesian inference, which are then clustered to reveal distinct behavioral types. 
C-Lasso \citep{su2016identifying, fallucchi2021identifying} extends this logic by simultaneously estimating and shrinking parameters in a single step.
At the data-driven end, unsupervised machine learning methods—such as clustering—identify behavioral patterns directly from the data \citep{Fallucchi2019, bolle2021behavioral}.

Theory-driven methods often impose assumptions about temporal dynamics—for example, that players respond not only to current but also to past group contributions. Such restrictions may limit their ability to capture evolving strategies, which could help explain why several studies report substantial unexplained variation, with residual clusters ranging from 16\% to 32\% of observations (\autoref{tab:behavioral_types}).
Even data-driven approaches face challenges. Despite their flexibility, existing clustering methods often rely on local distance metrics (e.g., Manhattan distance) that cannot account for misaligned or non-synchronous behavioral trajectories (details in App.~\ref{apx:methods}).

Returning to the earlier question: Could the combination of unaddressed temporal misalignment and the reliance on numerous assumptions contribute to the emergence of an ``other" cluster? 
How might we attempt to better understand such ``other" clusters? One promising direction lies in examining methods from the learning literature.

\paragraph{Types in Financial Decision Making}
Work in behavioral finance also relies heavily on time-series data to identify behavioral types.
The data contexts range from retail trading histories \citep{barber2001boys,kumar2009gamblers,dorn2005trading}, high-frequency order-flow and inventory trajectories \citep{kirilenko2017flash,brogaard2014high}, to longitudinal portfolio adjustments \citep{calvet2009financial}, fund return dynamics \citep{brown1996tournaments,wermers2000mutual}, and depositor withdrawal sequences during bank runs \citep{iyer2012understanding,trautmann2011bank,gorton2012misunderstanding}. 
A wide variety of behavioral types have been documented across these papers—including retail trading styles, high-frequency trading roles, mutual fund strategy types, and depositor responses in runs (see \autoref{tab:behavioral_finance}). 
However, this literature typically reduces behavioral trajectories to engineered features, summary statistics, or factor loadings.
As a result, interesting types of behavioral heterogeneity may remain hidden.

\paragraph{Reinforcement Learning vs. Intention Switching}
A long-standing question in behavioral game theory is whether reinforcement learning (RL), and specifically Q-learning, adequately describes how people behave in repeated social dilemmas. Early experimental work demonstrated that RL can capture aggregate patterns of play in repeated games such as the Prisoner’s Dilemma \citep{ErevRoth1998}. Hybrid models that combine RL with belief updating fit the data more accurately than either pure RL or belief learning alone \citep{CamererHo1999,HoCamererChong2007}. At the neural level, dopamine-based reward prediction error signals have been linked to RL processes, suggesting that Q-learning-like updating is plausible as a cognitive mechanism \citep{RangelCamererMontague2008}. Recent work continues to apply and extend Q-learning models to repeated social dilemmas, including the Prisoner’s Dilemma and public good games \citep{dolgopolov2024reinforcement,zheng2024evolution}.

These models primarily focus on how individuals update action values, rather than on the structure of the underlying reward function \citep{ErevRoth1998,CamererHo1999}. Yet specifying a reward function in a social dilemma is inherently complex: experimental and theoretical evidence shows that people incorporate social preferences—such as fairness, reciprocity, and inequality aversion—in addition to monetary payoffs \citep{FehrSchmidt1999,Rabin1993}. Moreover, even a high learning rate in an RL model describes continuous adaptation of beliefs and action values; it does not imply discrete switches in the underlying cooperative intention. Thus, while RL captures some aspects of behavioral adjustment in repeated interactions, it does not explain the sharp intention transitions that characterize the switcher type we identify.

\paragraph{Latent Variable Models}
Defining a comprehensive and suitable reward function poses challenges in complex behavioral tasks \citep{Alyahyay2023, Rosenberg2021}. 
Inverse Reinforcement Learning (IRL) \citep{Ng2000, Arora2021} is a popular approach to recover the reward function inductively from the data. IRL has achieved breakthrough successes in robotics \citep{Kumar2023,Chen2023} and autonomous driving \citep{Kalweit2020, Nasernejad2023} and has recently also become a valuable tool for constructing mathematical models of animal behavior \citep{Yamaguchi2018, Kwon2020, Alyahyay2023}.
Traditionally, the animal’s reward function has been modeled as a {\em smoothly} time-varying linear combination \citep{Ashwood2022a}.
More recently, it has been suggested that behavior can be represented as a Markov chain characterized by alternating between {\em discrete} intentions \citep{Ashwood2022b}. In that spirit, \citet{Zhu2024} propose a novel class of Hierarchical Inverse Q-Learning (HIQL) algorithms, which extend the fixed-reward inverse Q-learning (IQL) framework \citep{Kalweit2020} to solve multi-intention IRL problems.
Notably, HIQL achieves substantially better model fits to behavioral data compared to standard IQL, supporting the idea that accounting for discrete intentions improves the explanatory power of reward-based learning models.

\paragraph{This paper} 
This paper makes several contributions to the study of cooperation and learning in experimental settings. First, we introduce a method for identifying behavioral patterns in gameplay data that accommodates temporal misalignment—matching trajectories without requiring exact synchronization. We benchmark this DTW-based clustering approach against four established techniques: Bayesian modeling, finite mixture models, C-Lasso, and hierarchical clustering.
Second, we apply a latent variable model that allows for {\em discrete transitions} between latent intentions over time. This framework captures established behavioral types (free riders, unconditional cooperators) while providing a principled interpretation of the previously ambiguous ``other" cluster as highly exploratory players who systematically switch between strategies.

\section{Methods}
\label{sec:method}

The code used for this study will be available in a GitHub repository, which will be made public upon acceptance.

%\url{https://github.com/carinahausladen/IRL-public-good-games}.}. 
%To maintain anonymization, we have provided a link to the anonymized version:\url{https://anonymous.4open.science/r/IRL-public-good-games-1860/README.md}.

\subsection{The Linear Public Good Game} 

In each round of the linear public good game, every player receives an endowment \( e \) and decides how much to contribute \( c_{it} \) to a shared project. The total contribution is multiplied by a constant factor \( \mu \) and evenly distributed among all players. Each player’s payoff is $\pi_{it} = e - c_{it} + \mu \sum_{j=1}^I c_{jt}$, where \( 0 < \mu < 1 \) ensures that contributing benefits the group but is individually costly. To describe a player’s social environment, we define the average contribution of the other group members as $\bar{c}_{-i,t} = \frac{1}{I - 1} \sum_{\substack{j = 1 \\ j \neq i}}^I c_{jt}, \quad t = 1, 2, \dots, T.$

\subsection{Identifying Behavioral Types}

We propose a two-step approach to identify behavioral types. First, we cluster raw contribution trajectories using a global distance metric that measures similarity in temporal dynamics even when key moments (e.g., peaks or drops) occur at different rounds. This step treats the time series as the primitive object and remains fully model-free: it does not assume a specific objective function or imply that participants foresee the entire series when making decisions. Instead, the second step fits a latent-variable model to each cluster to provide behavioral structure by inferring the intention processes that best rationalize the observed patterns.

This approach adopts a finite-mixture perspective in which each trajectory is generated by one of a small number of latent behavioral types that systematically differ in how they respond over time. Our empirical-first, theory-second strategy follows the spirit of behavioral game theory and applied econometrics, which emphasize data-driven discovery followed by theoretically informed interpretation \citep{camerer2003models,mullainathan2017machine}.

\subsection{Clustering Multivariate Time-series} 
Clustering typically involves three steps: (1) selecting the number of clusters, (2) computing pairwise distances between time series, and (3) using these distances to form clusters.

\begin{figure}
    \centering
    \includegraphics[width=0.99\linewidth]{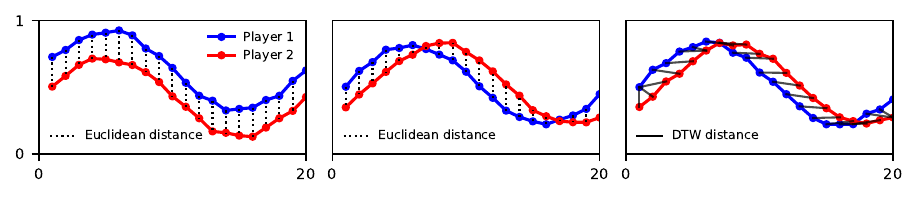}
    \caption{{\bf Conceptual Illustration of Euclidean and DTW distances.}(Left) Stacked sinusoidal curves represent two players with similar but vertically offset contribution patterns. (Center) Horizontally shifted curves illustrate temporal misalignment between players’ actions; dotted vertical lines denote pairwise Euclidean distances. (Right) DTW alignment paths (black solid lines) show the optimal nonlinear correspondence between the two time series, compensating for temporal shifts.}
    \label{fig:distance_metrics_demo}
\end{figure}

% TIME SHIFTED
In social dilemma experiments, participants are randomly assigned to groups and thus experience different dynamics—some groups sustain cooperation longer than others. Capturing such variation requires careful choice of the {\em distance measure} in step (2).

We illustrate the importance of this step using two stylized sinusoidal contribution patterns. In Fig.~\ref{fig:distance_metrics_demo} (left), the two curves are perfectly phase-aligned. In this case, the distance between the two time series can be effectively captured using a local distance metric, such as the point-wise Euclidean distance.
However, when the curves are phase-shifted (Fig.~\ref{fig:distance_metrics_demo}, middle), the Euclidean distance is no longer an effective similarity measure.
Behaviorally, such a phase shift could indicate that two participants are of similar behavioural types but belong to different experimental groups. In this case, a Euclidean comparison fails to group these participants.
DTW resolves this limitation by allowing non-linear alignments along the time axis (Fig.~\ref{fig:distance_metrics_demo}, right). Instead of comparing points strictly by time index, DTW aligns key structural moments—such as peaks—by warping the time dimension to minimize the overall alignment cost \citep[see][for details]{Berndt1994} (details in App.~\ref{apx:methods}).

%MULTIVARIATE
In addition to time-shifted behavior, social dilemma games have another key feature: they produce {\em multivariate time series}. 
In our data, participants play a standard PGG with feedback after each round, showing them the average contribution of their group members. Consequently, each participant is associated with two time series: their own contributions, $\{c_{it}\}_{t=1}^T$, and the average contributions of the other group members, $\{\bar{c}_{-i,t}\}_{t=1}^T$. 
DTW naturally accommodates such multivariate input, allowing us to compare behavioral trajectories across both dimensions (details in App.~\ref{apx:methods}).

After computing distances between multivariate time series, a {\em clustering algorithm} groups the data into meaningful clusters. 
We evaluate three clustering approaches—partition-based, hierarchical, and graph-based—applied to time series data using DTW as the distance metric. 
As is standard in unsupervised learning, we compare multiple algorithms and assess their stability and interpretability \citep{von2012clustering}. 
Our results show that the specific choice of clustering algorithm has little impact on the resulting cluster structure.

%\textbf{The Number of Clusters}
The optimal {\em number of clusters} $k$ is usually determined using internal cluster validation indices (CVIs). These indices assess clustering quality through different combinations of cluster cohesion and separation metrics \citep{Arbelaitz2013}. We employ three standard CVIs compatible with DTW-based clustering. We compute their average (normalized) score across candidate cluster numbers (2-20) to select the best number (App.~\ref{apx:n_clusters}). 

%\textbf{Centroid}
To interpret the clusters, it is common to compute a representative time series for each group. While simple averaging is often used, it fails when time series are misaligned. Instead, we use Dynamic Time Warping Barycenter Averaging (DBA) \citep{petitjean2011global}, which produces a centroid that preserves the overall shape and timing of behaviors within each cluster. To convey within-cluster heterogeneity, we display interquartile range (IQR) bands at each period.

%This i should transfer somewhere to the apx ideally at the interpretation of these plots
%the shaded region shows the 25th to 75th percentile of observed actions and states across all individuals in the cluster at that time point. These bands represent the empirical spread of behaviors, not sampling uncertainty. Classical confidence intervals are not appropriate for DBA barycenters because DBA is a nonlinear Fréchet mean defined in warped time, where alignment varies across individuals. Accordingly, DBA curves should be interpreted in conjunction with both the IQR bands and the underlying cluster heatmaps.

% Assessing Robustness
We assess robustness in three ways.
First, we examine sensitivity to modeling choices. While the choice of distance metric has a noticeable impact, different clustering algorithms (given our preferred 2-dimensional DTW metric) yield highly consistent partitions, indicating algorithmic robustness.
Second, we evaluate stability using a non-parametric bootstrap. In 100 replications, we resample 80\% of individuals, recompute DTW distances, re-estimate spectral clustering, and compare each solution to the baseline using ARI and NMI. This tests how strongly the structure persists under sampling variation.
Third, we analyze stability at the individual level. For each subject, we record their cluster assignment across bootstrap replications and compute the share of times they return to their baseline cluster.

\subsection{Latent Variable Model}
We apply a recently proposed latent variable model that has achieved 
superior model fit in modeling animal behavior as a Markov chain characterized by alternating between {\em discrete} intentions.
Specifically, this latent variable model is a Hierarchical Inverse Q-Learning (HIQL) algorithm \citep{Zhu2024}.

\paragraph{Q-learning.}
Q-learning is a fundamental algorithm in RL. The process is typically modeled as a Markov Decision Process (MDP), where the future state $s'$ depends only on the current state $s$ and the chosen action $a$, formally $P(s' \mid s, a)$. In the context of the PGG, each player's \emph{own contribution} can be treated as the \emph{action} $a \in \mathcal{A}$, while the \emph{observed contributions of others} form the \emph{state} $s \in \mathcal{S}$ \citep{zheng2024evolution}. 
Players are assumed to choose actions following an $\epsilon$-greedy policy, balancing exploration and exploitation:

\[
a =
\begin{cases}
\text{a random action}, & \text{with probability } \epsilon,\\[4pt]
\displaystyle \arg\max_a Q(s,a), & \text{with probability } 1 - \epsilon.
\end{cases}
\]

\begin{table}[]
\centering
\caption{Structure of a Q-table.}
\begin{tabular}{c|ccc}
\hline
\textbf{State} & $a_1$ & $a_2$ & $a_3$ \\
\hline
$s_1$ & $Q(s_1,a_1)$ & $Q(s_1,a_2)$ & $Q(s_1,a_3)$ \\
$s_2$ & $Q(s_2,a_1)$ & $Q(s_2,a_2)$ & $Q(s_2,a_3)$ \\
$s_3$ & $Q(s_3,a_1)$ & $Q(s_3,a_2)$ & $Q(s_3,a_3)$ \\
\hline
\end{tabular}

\label{tab:qlearning_pair}

\vspace{4pt}
\parbox{\linewidth}{\footnotesize\raggedright
\textit{Note:} Symbolic Q-values shown for each state–action pair.
}
\end{table}

Each player maintains a \emph{Q-table} (similar in spirit to Selten's (1967) Strategy Method, \autoref{tab:qlearning_pair}), where rows represent states and columns represent actions, and each entry $Q(s,a)$ encodes the estimated value of taking action $a$ in state $s$. Initially, the Q-table may contain random or neutral values, but these are updated over time using the standard Q-learning rule:

\begin{equation}
\resizebox{0.9\linewidth}{!}{$
Q_{\text{new}}(s,a)
=
(1 - \alpha)\, Q_{\text{old}}(s,a)
+
\alpha \left[ r + \gamma \max_{a'} Q_{\text{old}}(s',a') \right]
$}
\end{equation}

where $\alpha$ is the \emph{learning rate}, controlling how strongly new information replaces old estimates, $\gamma$ is the \emph{discount factor} that determines the importance of future rewards, and $r$ is the immediate reward received. 
Under the assumption of perfectly rational agents, the reward function corresponds directly to the payoff function of the PGG. However, empirical evidence suggests that human players exhibit social preferences and bounded rationality, making the reward function difficult to specify explicitly.
When this function is treated as unknown and instead inferred from observed behavior, the learning process falls within the class of {\em Inverse} Reinforcement Learning (IRL).

\paragraph{Inverse Q-learning.}
In \emph{inverse} Q-learning, the situation is reversed: we observe human behavior and aim to infer the underlying reward function that would make such behavior appear optimal. 
The human is assumed to behave like a Q-learner, but not a perfect one.
Given a dataset of observed game trajectories $\mathcal{D}$, the algorithm searches for a reward function $r(s,a)$ such that the resulting policy $\pi_r(s,a)$ would most likely have produced the same actions as the human. 

Mathematically, this can be expressed as maximizing the likelihood of the observed trajectories under the learned policy: $\max_{r} \; \mathbb{E}_{\xi \sim \mathcal{D}} \bigl[\log \Pr(\xi \mid \pi_r)\bigr]$,
where each trajectory $\xi = \{(s_0,a_0), \dots, (s_n,a_n)\}$ is a sequence of observed state–action pairs. The policy $\pi_r$ is derived from the learned action-value function $Q(s,a)$, which captures how good each action is in a given state. 
Inverse Q-learning thus provides a way to \emph{recover the hidden motivations} driving observed decisions, without assuming we already know how people evaluate outcomes (more details in \citep[][pp 3--5]{Zhu2024}).
The \emph{number of reward functions $r$}—or latent intentions—is treated as a model parameter and identified empirically through model fit criteria such as the log-likelihood and the Bayesian Information Criterion (BIC).

\paragraph{Hierarchical Extension.}
The hierarchical extension of the Inverse Q-learning (HIQL) algorithm makes an important conceptual leap. It assumes that observed human trajectories are generated by a two-level process:

At the lower level, \emph{intentions}—or latent reward functions $r$—determine the choice of actions $a$, which in turn govern the transition probabilities $P(s' \mid s,a)$ between states $s$. 
The transitions between intentions are modeled through a {\em discrete} Markov process with transition matrix $\Lambda$. 

\resizebox{0.8\linewidth}{!}{%
\begin{tikzpicture}[
  var/.style = {circle, draw, minimum size=25pt, inner sep=0pt},  % Fixed: all same size
  box/.style = {rectangle, draw, fill=gray!30, minimum size=8pt, inner sep=1pt},
  dots/.style = {font=\large},
  ->, >=Stealth, node distance=14mm
]
%--- coordinates
\coordinate (S1) at (-1.6, 1.0);
\coordinate (S2) at ( 0.0, 1.0);
\coordinate (S3) at ( 1.6, 1.0);
\coordinate (A1) at (-1.6, 0.0);
\coordinate (A2) at ( 0.0, 0.0);
\coordinate (R1) at (-1.6,-1.0);
\coordinate (R2) at ( 0.0,-1.0);
\coordinate (R3) at ( 1.6,-1.0);

%--- variables
\node[var] (st1) at (S1) {$s_{t-1}$};
\node[var] (st)  at (S2) {$s_t$};
\node[var] (stp) at (S3) {$s_{t+1}$};
\node[var] (at1) at (A1) {$a_{t-1}$};
\node[var] (at)  at (A2) {$a_t$};
\node[var] (rt1) at (R1) {$r_{t-1}$};
\node[var] (rt)  at (R2) {$r_t$};
\node[var] (rtp) at (R3) {$r_{t+1}$};

%--- transitions
\draw (st1) -- node[box, midway] (p1) {P} (st);
\draw (st)  -- node[box, midway] (p2) {P} (stp);
\draw (at1) -> (p1);
\draw (at)  -> (p2);
\draw (rt1) -- node[box, midway] {$\Lambda$} (rt);
\draw (rt)  -- node[box, midway] {$\Lambda$} (rtp);
\draw (rt1) -> (at1);
\draw (rt)  -> (at);

%--- horizontal inlet/outlet ellipses (ONLY top and bottom, NO middle)
\node[dots] (ldots_top) at (-2.8, 1.0) {$\cdots$};
\node[dots] (rdots_top) at ( 2.8, 1.0) {$\cdots$};
\node[dots] (ldots_bot) at (-2.8,-1.0) {$\cdots$};
\node[dots] (rdots_bot) at ( 2.8,-1.0) {$\cdots$};

%--- longer arrows with correct directions
\draw[->] (ldots_top) -- ++(0.6,0);
%\draw[->] (stp) -- ++(0.6,0);
\draw[->] (ldots_bot) -- ++(0.6,0);
\draw[<-] (rdots_top) -- ++(-0.6,0);  % Fixed: arrow points left
\draw[<-] (rdots_bot) -- ++(-0.6,0);  % Fixed: arrow points left

\end{tikzpicture}%
}

This is thus in contrast to previous approaches that assume \textit{continuous} shifts in latent rewards. 
The active intention at any given time is unobserved. The model therefore does not output a fixed reward function to each individual but instead estimates \emph{intention adoption probabilities}—the likelihood that a given behavior at time $t$ was generated under each latent intention. 

Notably, the shift from continuous to discrete transitions has improved model fit in recent studies of animal behavior. For our PGG data, this perspective may offer a potential explanation for the so-called ``other'' cluster, which often exhibits abrupt round-to-round changes in contribution patterns.

%Because the softmax decision rule smooths action probabilities, different reward parameters can produce nearly identical behaviors, leading to multiple observationally equivalent explanations of the same data. In this way, HIQL recovers the probabilistic landscape of underlying intentions rather than a single, fixed set of preferences.

\section{Data}
We use data gathered from standard linear public good games across different game length horizons (7, 10, 20, and 30 rounds). 
All except the 20-round subsets contain games without experimental intervention (like a punishment option, or a chat protocol). 
Furthermore, we only choose studies where the information treatment is such that players observe the {\em average contribution} of other team members after each round. 
Our dataset comprises 50,390 observations from 2,938 participants.
\autoref{tab:data_details} provides details of the original studies and several focal parameters, such as group size, number of rounds, and participants per study.
Fig.~\ref{fig:data} visualizes the raw data as a two-dimensional time series. 

\begin{table}
\centering
\caption{First-Round Cooperation Types by Period Length.}
\resizebox{\linewidth}{!}{%
\begin{tabular}{@{}lcccc|cc@{}}
\hline
Type $|$ Periods & 7 & 10 & 20 & 30 & \textbf{Ours} & T\&V 2018 \\
\hline
Free Rider (\%)             & 14 & 16 & 23 & 9  & \textbf{14.0} & 19.2 \\
Full Contributor (\%)       & 12 & 39 & 26 & 43 & \textbf{27.6} & 10.4 \\
Conditional Cooperator (\%) & 75 & 45 & 51 & 49 & \textbf{58.4} & 61.3 \\
\hline
\end{tabular}%
}
\vspace{4pt}
\parbox{\linewidth}{\footnotesize\raggedright
\textit{Note:} Percentages based on first-round normalized contributions in our pooled dataset. Our dataset comprises 50,390 observations from 2,938 participants. Benchmark values from \citet{thoni2018conditional}, based on 7,107 conditional-contribution observations (strategy method).
}
\label{tab:type_comparison}
\end{table}

To assess the representativeness of our dataset, we classify participants based on their first-round contributions (\autoref{tab:type_comparison}). Following \citet{cotla2019social}, we use initial behavior as a proxy for social preference types\footnote{First-round contributions strongly correlate with unconditional strategy-method classifications \citep{muller2008strategic,cotla2019social}.}. Using the thresholds from \citet{thoni2018conditional} — Free Riders ($c \leq 0.1$), Conditional Cooperators ($0.1 < c < 0.9$), and Full Cooperators ($c \geq 0.9$) — we find that Free Riders account for $14.0\%$ (vs.\ $19.2\%$ in the meta study), Conditional Cooperators for $58.4\%$ (vs.\ $61.3\%$), and Full Cooperators for $27.6\%$ (vs.\ $10.4\%$). While our dataset features more Full Cooperators and slightly fewer Conditional Cooperators, Conditional Cooperators remain the majority, indicating that our sample aligns well with canonical behavioral patterns.

\section{Results}
\label{sec:choice}

%The data subsets vary substantially in their experimental designs: group sizes range from 4 participants (standard in the literature) to several hundred (in the 7-round subset), the 20-round subset uniquely includes punishment treatments, and the 30-round games approach infinite-horizon conditions. 

Within the full dataset, the 10-round games with four-person groups constitute the modal experimental design \citep{ledyard1995handbook, chaudhuri2011sustaining}. Accordingly, Sections~\ref{sec:results_dtw} and~\ref{sec:results_intent} are based on this 10-round subset only.

\subsection{DTW-based Clustering Reveals Six Behavioral Clusters}
\label{sec:results_dtw}

We apply unsupervised learning to partition the multivariate time series data. The procedure consists of three steps: (1) selecting the {\em number of clusters}, (2) {\em computing distances} between time series, and (3) using these distances to {\em construct clusters}. 
For the 10-round subset, the optimal number of clusters is determined to be six (see App.~\ref{apx:n_clusters}). We adopt the same number of clusters when partitioning the remaining subsets (7, 20, and 30 rounds).

\begin{figure}
    \centering
    \includegraphics[width=0.99\linewidth]{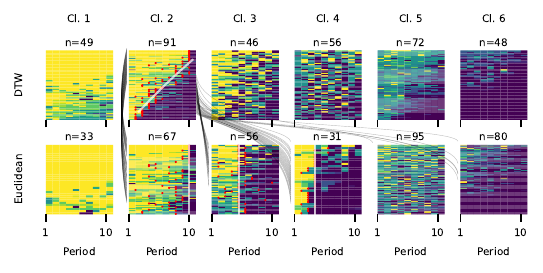}
    \caption{{\bf Global vs. Local Alignment.}Participants are clustered based on two different distance metrics: DTW (global) and Euclidean (local) distance.
Each small panel represents one cluster within a method, with participants sorted by their average contribution (highest on top).
Red dots indicate when the most substantial consecutive drop in contribution occurs per participant.
In the DTW cluster 2, a quadratic regression curve is overlaid to illustrate that the drop can occur at any point in the game.
In the Euclidean clusters 2, 3, and 4, a vertical reference line is drawn to mark the period where most detected decline points are concentrated.
Cell color encodes contribution magnitude (yellow indicates higher relative contribution, dark blue lower),
and sample sizes per cluster appear above each panel.}
    \label{fig:distance_metric_comparison}
\end{figure}

\paragraph{Global Alignment Outperforms Local Alignment.}
To illustrate the value of accommodating temporal misalignment when {\em computing distances}, we compare clustering results based on either local alignment (Euclidean distance) or global alignment (DTW; Fig.~\ref{fig:distance_metric_comparison}). DTW identifies a salient group (Cluster 2) characterized by a sharp transition from high to low contributions at different points in time. Under Euclidean distance, these participants are fragmented across multiple clusters based solely on when the switch occurs. In other words, Euclidean distance imposes artificial time-locking, whereas DTW correctly groups the shared strategic pattern of a qualitative shift from cooperation to defection (details in App.~\ref{apx:euclidean_vs_dtw}).

\begin{figure*}[!htpb]
    \centering
    \includegraphics[width=0.65\textwidth]{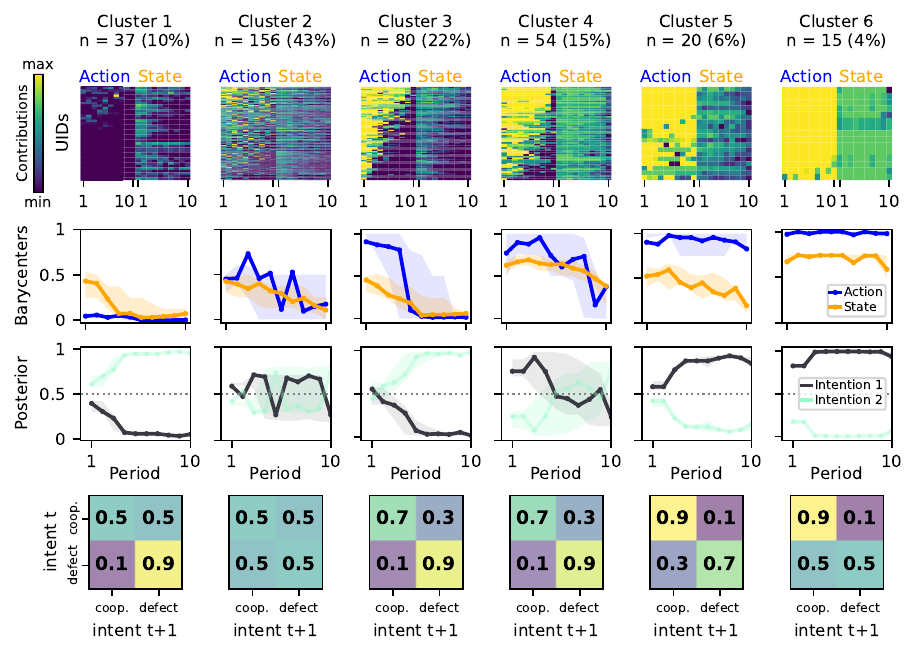}
    \caption{{\bf Clustering Analysis of Action-State Patterns and Intention Dynamics.} (1) Heatmaps of action and state trajectories by cluster, ordered by mean action. Clustering uses two-dimensional time series of own contributions (action) and others’ average contributions (state), normalized to [0,1]. Rows denote individuals (UIDs), and rows as well as clusters are ordered from lowest to highest average contribution.
(2) DTW barycenters of action-state patterns per cluster, with interquartile ranges shaded.
(3) DTW barycenter averages of posterior intention adoption probabilities by cluster, with interquartile ranges shaded. Two intentions are estimated; the second is the complement of the first. Values closer to one indicate a higher likelihood that the intention is adopted.
(5) Average transition probabilities between cooperative and defecting intentions per cluster.}
    \label{fig:master_panel_10}
\end{figure*}

Once DTW is selected as the appropriate similarity metric, varying the clustering algorithm results in only minor differences in the partition (App.~\ref{apx:cluster_algorithms}). We therefore adopt spectral clustering to {\em construct clusters}. 
The clustered raw data can be seen in the first row in Fig. \ref{fig:master_panel_10}.
Finally, compared to established methods such as hierarchical clustering \citep{Fallucchi2019}, C-Lasso \citep{su2016identifying}, and finite mixture models \citep{bardsley2007experimetrics}, our approach produces the cleanest and most interpretable segmentation of behavioral types (Fig.~\ref{fig:method_comp}).

These analyses confirm that the identified clusters are not artifacts of specific modeling choices. We next assess whether they are also stable with respect to sampling variation.

\paragraph{Clusters Are Robust to Sampling Variation Yet Retain Within-Cluster Heterogeneity.}
We assess cluster quality using bootstrap resampling (100 replications) and heterogeneity metrics.
The clustering solution is highly stable: bootstrap partitions closely match the original assignment (ARI = 0.861, SD = 0.061; NMI = 0.868, SD = 0.047), and participants return to their dominant cluster in 29.9\% of replications (SD = 3.3 pp), well above the 16.7\% expected under random assignment (\autoref{tab:cluster_diag_sidebyside}).

At the same time, clusters intentionally retain substantial internal heterogeneity.
Although all trajectories are assigned to one of six clusters, clusters explain only 29.5\% of total DTW-based trajectory variance, leaving 70.5\% within clusters.
Individual fit varies widely: 26.9\% of participants are closer to an alternative cluster than their assigned cluster (silhouette $<0$), and between-cluster separation is modest (ratio $=1.56$; see \autoref{tab:cluster_diag_sidebyside}).

Taken together, cluster assignments are robust to resampling yet leave ample room for the idiosyncratic and stochastic variation in individual contribution behavior that is well documented in the PGG literature.

\subsection{Two Intentions Successfully Integrate Actions and States}
\label{sec:results_intent}
%The clustering results on the 10-round subset are presented in \autoref{fig:master_panel_10}, row 1. Visualizations of all subsets and their corresponding clusters can be found in \autoref{fig:all_heatmaps_actions_states}.

To interpret the clusters, we estimate a latent variable model \citep{Zhu2024} based on the observed player contributions (actions) and the average contributions of their group members (states).

We find that increasing the number of latent intentions from one to two yields the largest improvement in test log-likelihood (LL), accompanied by the smallest increase in BIC ($\Delta$ LL = 0.66; $\Delta$ BIC = 75.2, \autoref{tab:likelihood_bic}). 
We thus assume two latent intentions.
Model fit on two latents across clusters is detailed in \autoref{tab:cluster_loglikelihood}, column 2.

\paragraph{Intentions Adoption Probabilities Allow for the Joint Interpretation of States and Actions.}

The latent-variable model outputs posterior intention adoption probabilities. Because this quantity is latent, it must be interpreted in relation to the observed player contributions (actions) and group average contributions (states).
This relationship becomes clear when examining individual examples in Fig.~\ref{fig:samples}, where intention probabilities are plotted alongside actions and states. In some cases, intentions closely track behavioral choices; in others, they align more strongly with state dynamics. 
In other words, the intention adoption probabilities formally integrate both observable data—actions and states—into a single representational trajectory. 
This integration allows researchers to interpret participants’ underlying decision tendencies without relying on ad hoc comparisons between separate state and action patterns.
%Consequently, intention adoption probabilities offer a more principled and interpretable way to characterize each behavioral cluster and to understand strategic decision-making in the PGG.

\paragraph{Latent Intentions Successfully Model Stable and Switching Decision Patterns.}

We aggregate individual posterior trajectories within each cluster using Barycenter averaging (Fig.~\ref{fig:master_panel_10}, row 3). Clusters 1, 5, and 6 display well-separated intention-adoption probabilities that change little after early rounds. 
This suggests that participants in these clusters form stable strategic intentions and adhere to them throughout the game. 

In the behavioral data (Fig.~\ref{fig:master_panel_10}, rows 1 and 2), these intentions correspond to persistent free-riding (Cluster 1) and sustained cooperation (Clusters 5–6). Cluster 1 exhibits uniformly low contributions, visible as a dark blue heatmap and a flat DTW barycenter near zero. In contrast, Cluster 6 shows consistently high contributions, with a near-uniform yellow heatmap and a barycenter that remains close to one throughout.

Because the model captures these established types, we can also interpret Cluster 2 with confidence. This group exhibits frequent alternation between cooperation and defection, often from one round to the next. Their inferred transition probabilities (Fig.~\ref{fig:master_panel_10}, row 4) show a roughly 50\% chance of switching intentions each period—indicating neither noise nor gradual learning, but a distinct decision-making mode characterized by continuous strategy experimentation. We therefore refer to this behavioral type as ``Switchers": players who actively and frequently cycle between intentions.

The discrete latent-intention formulation thus provides a unified framework that accommodates both stable and erratic behavioral dynamics within the same generative process. 
It confirms long-known strategic types and additionally suggests that high-frequency intention switching constitutes a meaningful and interpretable behavioral pattern in its own right.

\subsection{Interpretation of the Clusters}

Overall, out of the six behavioral patterns, two maintain stable strategies. Cluster~1 represents persistent \emph{free riders} who defect throughout despite favorable early outcomes, while Cluster~6 comprises \emph{consistent cooperators} who sustain high contributions when experiencing positive feedback. Cluster~5 are \emph{unconditional cooperators} who maintain high contributions even under unfavorable conditions. Note that the distinction between Clusters~5 and~6 becomes identifiable through DTW’s two-dimensional representation, which incorporates both an individual’s contributions and the surrounding group behavior. Our approach identifies these types across game lengths—free riders and consistent cooperators appear in all subsets, while unconditional cooperators emerge only in the shorter horizons. 
Please consult Fig.~\ref{fig:type-comparison} for a clear visual comparison of types/clusters across game lengths.

The remaining clusters reveal dynamic forms of cooperation breakdown. 
Cluster~3 consists of \emph{threshold switchers} who begin with near-maximal cooperation but then abruptly transition to persistent free riding. This discontinuity is clearly visible in both the raw data (row 1) and barycenters (Fig.~\ref{fig:master_panel_10}, row 2). The pattern resembles classical discontinuous adjustment mechanisms: \emph{grim-trigger} strategies that cooperate until a perceived deviation and then punish permanently \citep{friedman1971supergames, axelrod1984evolution}, Win–Stay Lose–Shift rules that switch when outcomes fall below aspiration \citep{nowak1993pavlov, macy2002learning}, and adaptive-aspiration models generating endogenous switching thresholds \citep{karandikar1998aspirations}. However, unlike planned endgame defection, the switch in Cluster~3 does occur at any round, suggesting a reactive breakdown rather than horizon-based planning. This type is most clearly observed in the 10-round subset but also appears more weakly in the 30-round games.

Cluster~4 exhibits a noisier decline: players start with high cooperation but reduce contributions gradually, with full defection concentrated toward the final rounds. 
Patterns of high early cooperation followed by a sharp collapse near the end of the game are broadly consistent with the classic ``endgame effect'' in finitely repeated social dilemmas \citep[e.g.][]{keser2000conditional,gaechter2007conditional,embrey2017cooperation}. In public-goods experiments, for instance, \citet{gaechter2005social} document groups that maintain substantial contributions up to the penultimate round and then free ride in the final period, and interpret this as rational last-round defection. \citet{engelrockenbach2020} and \citet{engel2024precarious} interpret this behavior as \emph{farsighted free-riders} who ``feed the cow'' by contributing in early periods and then ``milk'' it by cutting contributions once exploitation becomes profitable. Our Cluster~4 matches this pattern but is identified in a fully data-driven way.

Finally, \textit{Cluster~2} appears erratic at first glance, but becomes interpretable under a latent-state perspective as {\em Switchers}. We examine this behavior in detail in the next subsection.

\subsection{``Switchers" across subsets}

\begin{figure}
    \centering
    \includegraphics[width=0.99\linewidth]{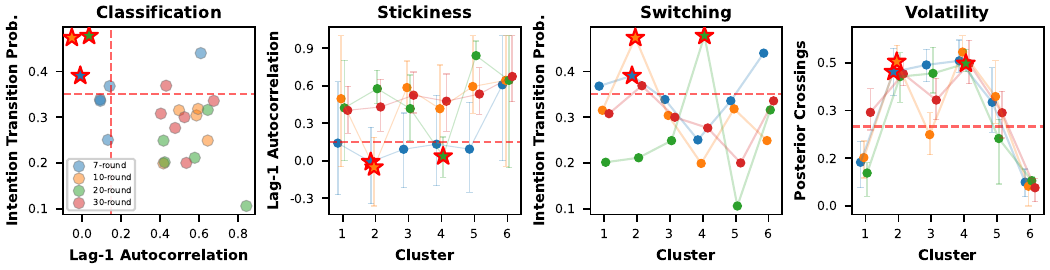}
    \caption{{\bf High-Switcher Identification Through Multi-Metric Behavioral Classification.} (\textbf{A}) Classification of all 24 behavioral clusters across four game lengths (7, 10, 20, 30 rounds) in a two-dimensional space defined by \emph{stickiness} (lag-1 autocorrelation) and \emph{switching rate} (mean transition probability between latent cooperative and defective intentions). Switchers (stars) meet all three criteria. Dashed red lines show decision boundaries. 
(\textbf{B--D}) Cluster-level metrics with interquartile ranges (error bars).
(\textbf{B}) Switchers exhibit near-zero stickiness.  
(\textbf{C}) Switching rates for Switchers (39--48\%) approach random-walk behavior ($50\%$), indicating frequent reversals in latent intention.  
(\textbf{D}) Posterior volatility reflects how often inferred intentions cross the $0.5$ threshold. Only Switchers pair this with low stickiness and high switching.}
    \label{fig:alternator}
\end{figure}

To quantify the behavior of \textit{Cluster~2} in Fig. \ref{fig:master_panel_10}, we define three properties of intention dynamics: \textit{stickiness} (lag-1 autocorrelation), \textit{switching rate} (intention transition probabilities), and \textit{volatility} (posterior threshold crossings). 

We define ``Switchers" using three empirically-motivated thresholds: (1) \textit{stickiness} $< 0.15$ (conventional threshold for weak temporal dependence), (2) \textit{switching rate} $> 0.35$ (75th percentile, representing elevated but sub-random transition rates), and (3) \textit{volatility} $> 0.25$ (indicating posterior reversals approximately every four rounds). 

It is important to note that all three metrics need to be present to define the behavior of Cluster 2 because each metric alone fails to discriminate this behavior reliably. 
%\textit{Stickiness} is confounded by short game horizons and does not capture intentional directionality. The \textit{switching rate} can appear artificially high for stable cooperators who occasionally correct minor deviations. \textit{Volatility} alone is broadly elevated outside extreme clusters, making it insufficient as a standalone criterion. Only the combination of all three features isolates a coherent and recurrent behavioral profile.

Across datasets, {\em Switchers} emerge as a nontrivial share of participants. They appear in three of four game-length subsets: $456$ participants in the 7-round subset ($41 \%$), $123$ participants in the 10-round subset ($34 \%$), and $49$ participants in the 20-round subset ($10 \%$). No clusters in the 30-round subset meet all three criteria. In total, Switchers represent $628$ out of $N = 2938$ participants ($21.4 \%$). Their grouping via DTW-based trajectory similarity confirms they form a distinct cluster rather than residual outliers forced together. 
%Visual inspection shows rapid oscillation without temporal trends, in contrast to other clusters that exhibit clear movement toward cooperation or defection.

{\em Switchers} show near-zero \textit{stickiness}, \textit{switching rate} rates close to $50 \%$, and weak behavioral responsiveness to peers' contributions ($r = 0.14$, compared with $r = 0.41$ in other types). 
The prevalence of {\em Switchers} declines systematically with game length: $41 \%$ in 7-round games, $34 \%$ in 10-round games, $10 \%$ in 20-round games, and $0 \%$ in 30-round games. 
This pattern may suggest early-stage strategic uncertainty that resolves as experience accumulates: In short interactions, many participants remain uncommitted and explore both strategic directions. When more time is available, they tend to converge toward stable cooperative or defective patterns.

\section{Discussion}
Our main contributions are twofold. 
First, we show that pattern-based alignment using Dynamic Time Warping (DTW) substantially improves behavioral clustering in repeated-game environments relative to conventional Euclidean alignment. By grouping trajectories according to their temporal shape rather than round-by-round proximity, we uncover structure in dynamic behavior that is both interpretable and theoretically meaningful. 
Second, we apply a latent‐variable framework in which individuals switch \emph{discretely} between intentions over time. By formally integrating the two observable time series, our approach provides a principled interpretation of each cluster. This approach recovers well-known behavioral types (such as persistent free riders and unconditional cooperators) that have been repeatedly validated in the literature, while also delivering the first rigorous account of the previously ambiguous ``other" group.

\paragraph{A Unifying Framework.}
Our two-step approach provides a {\em unifying framework} for explaining \emph{all} observed behavioral clusters in the PGG literature. Prior work typically imposes theory-driven assumptions that successfully capture a few well-established types, but then relegate the remaining behavior to error terms or ``unexplained heterogeneity.'' In contrast, every cluster we uncover is interpretable through the latent-variable structure: even the previously ambiguous ``other'' group receives a coherent interpretation via three key parameters—autocorrelation, transition probabilities, and intention crossings. 

This unified perspective yields two particularly substantive insights for research on cooperation dynamics. 
First, we identify a clear class of \emph{threshold switchers} who begin with near-maximal cooperation before abruptly transitioning to persistent free riding. Prior work has inferred such strategies either by restricting subjects to a pre-specified menu of strategies \citep[e.g.][]{dal2019strategy}, or by estimating behavior within narrow parametric classes such as memory-one HMMs \citep{montero2022inferring} or finite-state automata \citep{kleiman2018non}. By contrast, our identification operates directly on unconstrained contribution trajectories and recovers this cluster without any prior specification of memory length, state space, or strategy catalog. To our knowledge, this is the first purely data-driven recovery of a grim-trigger-like pattern from human play.
Second, we show that a substantial fraction of individuals exhibit \emph{volatile but recoverable} intention dynamics. It is important to interpret this behavior: Standard models treat defections as informative and persistent, triggering punishment and the collapse of cooperation. Our findings show that a substantial fraction of individuals exhibit volatile intention dynamics: deviations are frequent but reversible. In settings where cooperation underpins long-term value—supply chains, alliances, subscription services, customer retention, and online communities—misinterpreting such transient defections can destroy value by breaking {\em relationships that would have recovered}. Recognizing intentional volatility expands the strategy space: forward-looking partners may rationally exercise {\em forgiveness and strategic patience}, maintaining cooperation because they expect switching types to return to cooperative behavior.

\paragraph{Managerial Relevance.}
Our framework is fully data-driven and therefore well-suited for settings where behavior is complex and controlled laboratory experimentation is infeasible. We benchmark its validity in a mature and robust literature and find that it reliably recovers well-established types and yields plausible interpretations for all patterns found. In this sense, the PGG environment functions as a stress test: if the method can reconstruct and interpret a space of behaviors that is already well understood, it becomes a credible tool for domains where theory remains underdeveloped.
Behavioral finance provides a prime illustration. A wide variety of trading and portfolio archetypes has been documented—ranging from retail trader styles and high-frequency trading roles to mutual-fund strategies and depositor behavior in bank runs (see \autoref{tab:behavioral_finance}). However, existing approaches often compress rich time-series behavior into engineered features or factor exposures, obscuring temporal structure and strategic heterogeneity. Our approach is designed to thrive in exactly such environments. Financial decisions generate large panel time-series datasets, and our method operates directly on the \emph{full trajectories} rather than on prespecified summaries.

\paragraph{Straightforward Applicability.}
Our approach is designed for straightforward adoption. In practice, analysts only need to provide a panel of (potentially multidimensional) time series; the pipeline performs DTW-based clustering, followed by fitting the hierarchical inverse Q-learning.
A central challenge is visualizing rich behavioral trajectories. Standard line plots become unreadable when many individuals are shown, and existing heatmaps often collapse the time dimension, often plotting state against actions. We therefore introduce an effective visualization scheme for multidimensional time series: raw trajectories are displayed as heatmaps sorted by average contribution. The sorting, in particular, makes behavioral patterns immediately visible even when they differ in timing or absolute levels. DTW-based barycenters are plotted alongside, preserving characteristic temporal structure without the over-smoothing inherent in simple averages.
These elements make the workflow highly transferable. Practitioners can reuse our plotting tools, apply DTW-based spectral clustering to their own data, and interpret resulting clusters via the latent reward structure. All code will be made publicly available on GitHub.

\paragraph{Limitation.}
A natural limitation of our approach is that DTW clusters contribute sequences based on their observable temporal shape rather than on underlying intentions. This means that similar contribution paths may arise from different motives. In addition, DTW treats similar behaviors at different time points as comparable, an assumption that may be imperfect in finite-horizon games where the interpretation of actions can vary over time.
We partially address this concern in the second stage by applying the HIQL model, which incorporates forward-looking behavior through Q-learning and discount factors that capture horizon effects. While this does not fully resolve the issue, it allows us to interpret DTW-based behavioral regularities through a model that explicitly accounts for temporal decision structure.

\paragraph{Future Work.}
Our model adapts reinforcement-learning tools developed for animal navigation to human cooperation. A natural extension is to integrate human-specific drivers—such as aspirations and affective responses—so that latent intentions more fully capture social motivations. Methodologically, we estimate inverse Q-learning \emph{within} clusters; evaluating how results change without clustering, or with alternative segmentation, is an important next step. Allowing cluster-specific discount factors may also reveal differences in foresight, though lower values require additional smoothing to ensure convergence.
Several technical refinements could improve estimation: incorporating first-round actions directly into state inference, aligning intention initialization with clear behavioral types (e.g., unconditional cooperators), and assessing alternative binning strategies to retain richer behavioral detail. Together, these extensions would increase psychological realism, robustness, and scalability to larger applied settings.

{
\small

% Bibliography
%\section*{References}
\clearpage
\bibliographystyle{unsrtnat}
\bibliography{refs}

@article{abeler2015self,
    author          = {Abeler, Johannes and Nosenzo, Daniele},
    doi             = {10.1007/s10683-014-9397-9},
    journal         = {Experimental Economics},
    number          = {2},
    pages           = {195--214},
    publisher       = {Springer},
    title           = {Self-selection into laboratory experiments: Pro-social motives versus monetary incentives},
    volume          = {18},
    year            = {2015},
}

@article{aimone2013endogenous,
    author          = {Aimone, Jason A and Iannaccone, Laurence R and Makowsky, Michael D and Rubin, Jared},
    doi             = {10.1093/restud/rdt017},
    journal         = {The Review of Economic Studies},
    number          = {4},
    pages           = {1215--1236},
    publisher       = {Oxford University Press},
    title           = {Endogenous group formation via unproductive costs},
    volume          = {80},
    year            = {2013},
}

@article{Alyahyay2023,
    author          = {Alyahyay, Mansour and Kalweit, Gabriel and Kalweit, Maria and Karvat, Golan and Ammer, Julian and Schneider, Artur and Adzemovic, Ahmed and Vlachos, Andreas and Boedecker, Joschka and Diester, Ilka},
    doi             = {10.1101/2023.01.20.524944},
    journal         = {bioRxiv},
    note            = {Preprint},
    pages           = {2023--01},
    title           = {Mechanisms of premotor-motor cortex interactions during goal directed behavior},
    year            = {2023},
}

@article{Andreoni1995,
    author          = {Andreoni, James},
    journal         = {American Economic Review},
    number          = {4},
    pages           = {891--904},
    title           = {Cooperation in public goods experiments: Kindness or confusion},
    volume          = {85},
    year            = {1995},
}

@article{Arbelaitz2013,
    author          = {Arbelaitz, Olatz and Gurrutxaga, Ibai and Muguerza, Javier and P{\'e}rez, Jes{\'u}s M and Perona, I{\~n}igo},
    doi             = {10.1016/j.patcog.2012.07.021},
    journal         = {Pattern Recognition},
    number          = {1},
    pages           = {243--256},
    publisher       = {Elsevier},
    title           = {An Extensive Comparative Study of Cluster Validity Indices},
    volume          = {46},
    year            = {2013},
}

@article{Arora2021,
    author          = {Arora, Saurabh and Doshi, Prashant},
    doi             = {10.1016/j.artint.2021.103500},
    journal         = {Artificial Intelligence},
    pages           = {103500},
    title           = {A survey of inverse reinforcement learning: Challenges, methods and progress},
    volume          = {297},
    year            = {2021},
}

@article{Ashwood2022a,
    author          = {Ashwood, Zoe and Jha, Aditi and Pillow, Jonathan W},
    journal         = {Advances in Neural Information Processing Systems},
    pages           = {29663--29676},
    title           = {Dynamic inverse reinforcement learning for characterizing animal behavior},
    volume          = {35},
    year            = {2022},
}

@article{Ashwood2022b,
    author          = {Ashwood, Zoe C. and Roy, Nicholas A. and Stone, Iris R. and International Brain Laboratory and Urai, Anne E. and Churchland, Anne K. and Pouget, Alexandre and Pillow, Jonathan W.},
    doi             = {10.1101/2020.10.19.346353},
    journal         = {Nature Neuroscience},
    number          = {2},
    pages           = {201--212},
    title           = {Mice alternate between discrete strategies during perceptual decision-making},
    volume          = {25},
    year            = {2022},
}

@book{axelrod1984evolution,
    author          = {Axelrod, Robert},
    publisher       = {Basic Books},
    title           = {The Evolution of Cooperation},
    year            = {1984},
}

@article{barber2001boys,
    author          = {Barber, Brad M. and Odean, Terrance},
    doi             = {10.2139/ssrn.139415},
    journal         = {Quarterly Journal of Economics},
    number          = {1},
    pages           = {261--292},
    title           = {Boys will be boys: Gender, overconfidence, and common stock investment},
    volume          = {116},
    year            = {2001},
}

@article{bardsley2007experimetrics,
    author          = {Bardsley, Nicholas and Moffatt, Peter G},
    doi             = {10.1007/s11238-006-9013-3},
    journal         = {Theory and Decision},
    number          = {2},
    pages           = {161--193},
    publisher       = {Springer},
    title           = {The experimetrics of public goods: Inferring motivations from contributions},
    volume          = {62},
    year            = {2007},
}

@article{Bayer2013,
    author          = {R. C. Bayer and E. Renner and R. Sausgruber},
    doi             = {10.1007/s10683-012-9348-2},
    journal         = {Experimental Economics},
    number          = {4},
    pages           = {478--496},
    title           = {Confusion and learning in the voluntary contributions game},
    volume          = {16},
    year            = {2013},
}

@inproceedings{Berndt1994,
    address         = {Seattle, WA, USA},
    author          = {Berndt, Donald J. and Clifford, James},
    booktitle       = {Proceedings of the 3rd International Conference on Knowledge Discovery and Data Mining (KDD Workshop)},
    pages           = {359--370},
    title           = {Using Dynamic Time Warping to Find Patterns in Time Series},
    year            = {1994},
}

@article{bolle2021behavioral,
    author          = {Bolle, Friedel and Tan, Jonathan HW},
    doi             = {10.1007/s40881-021-00101-z},
    journal         = {Journal of the Economic Science Association},
    number          = {1},
    pages           = {49--63},
    publisher       = {Springer},
    title           = {Behavioral types of the dark side: identifying heterogeneous conflict strategies},
    volume          = {7},
    year            = {2021},
}

@unpublished{BordtFarbmacher2019,
    author          = {Sebastian Bordt and Helmut Farbmacher},
    note            = {Working Paper},
    title           = {Estimating Grouped Patterns of Heterogeneity in Repeated Public Goods Experiments},
    year            = {2019},
}

@article{BoschDomenech2010,
    author          = {Bosch-Domenech, Antoni and Montalvo, Jose G and Nagel, Rosemarie and Satorra, Albert},
    doi             = {10.1007/s10683-010-9251-7},
    journal         = {Experimental Economics},
    number          = {4},
    pages           = {461--475},
    title           = {A finite mixture analysis of beauty-contest data using generalized beta distributions},
    volume          = {13},
    year            = {2010},
}

@article{Brocas2014,
    author          = {Brocas, Isabelle and Carrillo, Juan D and Wang, Stephanie W and Camerer, Colin F},
    doi             = {10.1093/restud/rdu001},
    journal         = {Review of Economic Studies},
    number          = {3},
    pages           = {944--970},
    title           = {Imperfect choice or imperfect attention? Understanding strategic thinking in private information games},
    volume          = {81},
    year            = {2014},
}

@article{brogaard2014high,
    author          = {Brogaard, Jonathan and Hendershott, Terrence and Riordan, Ryan},
    journal         = {Review of Financial Studies},
    number          = {8},
    pages           = {2267--2306},
    title           = {High-Frequency Trading and Its Impact on Market Quality},
    volume          = {27},
    year            = {2014},
}

@article{brown1996tournaments,
    author          = {Brown, Keith C. and Harlow, W. Van and Starks, Laura T.},
    doi             = {10.2307/2329303},
    journal         = {Journal of Finance},
    number          = {1},
    pages           = {85--110},
    title           = {Of Tournaments and Temptations: An Analysis of Managerial Incentives in the Mutual Fund Industry},
    volume          = {51},
    year            = {1996},
}

@article{burlando2005heterogeneous,
    author          = {Burlando, Roberto and Guala, Francesco},
    doi             = {10.1007/s10683-005-0436-4},
    journal         = {Experimental Economics},
    number          = {1},
    pages           = {35--54},
    publisher       = {Springer},
    title           = {Heterogeneous agents in public goods experiments},
    volume          = {8},
    year            = {2005},
}

@article{burton2022strategy,
    author          = {Burton-Chellew, Maxwell N and D’Amico, Victoire and Gu{\'e}rin, Claire},
    doi             = {10.3390/g13060069},
    journal         = {Games},
    number          = {6},
    pages           = {69},
    publisher       = {MDPI},
    title           = {The Strategy Method Risks Conflating Confusion with a Social Preference for Conditional Cooperation in Public Goods Games},
    volume          = {13},
    year            = {2022},
}

@article{BurtonChellew2013,
    author          = {M. N. Burton-Chellew and S. A. West},
    journal         = {Proceedings of the National Academy of Sciences},
    number          = {6},
    pages           = {216--221},
    title           = {Prosocial preferences do not explain human cooperation in public goods games},
    volume          = {110},
    year            = {2013},
}

@article{BurtonChellew2016,
    author          = {M. N. Burton-Chellew and C. El Mouden and S. A. West},
    journal         = {Proceedings of the National Academy of Sciences},
    number          = {6},
    pages           = {1291--1296},
    title           = {Conditional cooperation and confusion in public goods experiments},
    volume          = {113},
    year            = {2016},
}

@article{calinski1974dendrite,
    author          = {Cali{\'n}ski, Tadeusz and Harabasz, Jerzy},
    journal         = {Communications in Statistics-theory and Methods},
    number          = {1},
    pages           = {1--27},
    publisher       = {Taylor \& Francis},
    title           = {A dendrite method for cluster analysis},
    volume          = {3},
    year            = {1974},
    doi             = {10.1080/03610917408548446},
}

@article{calvet2009financial,
    author          = {Calvet, Laurent E and Campbell, John Y and Sodini, Paolo},
    doi             = {10.3386/w14699},
    journal         = {American Economic Review},
    number          = {2},
    pages           = {393--398},
    title           = {Measuring the Financial Sophistication of Households},
    volume          = {99},
    year            = {2009},
}

@article{camerer2003models,
    author          = {Camerer, Colin and Ho, Teck and Chong, Kuan},
    doi             = {10.1257/000282803321947038},
    journal         = {American Economic Review},
    number          = {2},
    pages           = {192--195},
    publisher       = {American Economic Association},
    title           = {Models of thinking, learning, and teaching in games},
    volume          = {93},
    year            = {2003},
}

@article{Camerer2006,
    author          = {C. F. Camerer and E. Fehr},
    journal         = {Science},
    number          = {5757},
    pages           = {47--52},
    title           = {When does ``economic man'' dominate social behavior?},
    volume          = {311},
    year            = {2006},
}

@article{CamererHo1999,
    author          = {Camerer, Colin and Ho, Teck-Hua},
    doi             = {10.1111/1468-0262.00054},
    journal         = {Econometrica},
    number          = {4},
    pages           = {827--874},
    title           = {Experience-weighted attraction learning in normal form games},
    volume          = {67},
    year            = {1999},
}

@article{Chaudhuri2011,
    author          = {Chaudhuri, Ananish},
    doi             = {10.1007/s10683-010-9257-1},
    journal         = {Experimental Economics},
    number          = {1},
    pages           = {47--83},
    publisher       = {Springer},
    title           = {Sustaining Cooperation in Laboratory Public Goods Experiments: a Selective Survey of the Literature},
    volume          = {14},
    year            = {2011},
}

@article{chaudhuri2011sustaining,
    author          = {Chaudhuri, Ananish},
    journal         = {Experimental Economics},
    number          = {1},
    pages           = {47--83},
    publisher       = {Springer},
    title           = {Sustaining cooperation in laboratory public goods experiments: a selective survey of the literature},
    volume          = {14},
    year            = {2011},
    doi             = {10.1007/s10683-010-9257-1},
}

@inproceedings{Chen2023,
    author          = {Chen, Jiayu and Lan, Tian and Aggarwal, Vaneet},
    booktitle       = {2023 IEEE International Conference on Robotics and Automation (ICRA)},
    doi             = {10.1109/icra48891.2023.10160374},
    organization    = {IEEE},
    pages           = {5902--5908},
    title           = {Option-aware adversarial inverse reinforcement learning for robotic control},
    year            = {2023},
}

@article{Conte2011,
    author          = {Conte, Anna and Hey, John D and Moffatt, Peter G},
    doi             = {10.1142/9789813235816_0001},
    journal         = {Journal of Econometrics},
    number          = {1},
    pages           = {79--88},
    title           = {Mixture models of choice under risk},
    volume          = {162},
    year            = {2011},
}

@unpublished{cotla2019social,
    author          = {Cotla, Chenna Reddy and Petrie, Ragan},
    note            = {Working paper},
    title           = {Social Preferences and Payoff-Based Learning Explain Contributions in Repeated Public Goods Games},
    url             = {http://www.raganpetrie.org/uploads/8/4/4/3/84436206/prefs_and_learning_paper_nov2019.pdf},
    year            = {2019},
}

@article{cubitt2017conditional,
    author          = {Cubitt, Robin and G{\"a}chter, Simon and Quercia, Simone},
    doi             = {10.1016/j.jebo.2017.06.013},
    journal         = {Journal of Economic Behavior \& Organization},
    pages           = {110--121},
    publisher       = {Elsevier},
    title           = {Conditional cooperation and betrayal aversion},
    volume          = {141},
    year            = {2017},
}

@article{dal2019strategy,
    author          = {Dal B{\'o}, Pedro and Fr{\'e}chette, Guillaume R},
    journal         = {American Economic Review},
    number          = {11},
    pages           = {3929--3952},
    publisher       = {American Economic Association 2014 Broadway, Suite 305, Nashville, TN 37203},
    title           = {Strategy choice in the infinitely repeated prisoner’s dilemma},
    volume          = {109},
    year            = {2019},
    doi             = {10.1257/aer.20181480},
}

@article{dariel2014cooperators,
    author          = {Dariel, Alexis and Nikiforakis, Nikos},
    doi             = {10.1016/j.econlet.2013.10.033},
    journal         = {Economics Letters},
    number          = {2},
    pages           = {163--166},
    publisher       = {Elsevier},
    title           = {Cooperators and reciprocators: A within-subject analysis of pro-social behavior},
    volume          = {122},
    year            = {2014},
}

@article{Diederich2016,
    author          = {Diederich, Johannes and Goeschl, Timo and Waichman, Israel},
    doi             = {10.1016/j.euroecorev.2016.03.001},
    journal         = {European Economic Review},
    pages           = {272--287},
    publisher       = {Elsevier},
    title           = {Group Size and the (In)Efficiency of Pure Public Good Provision},
    volume          = {85},
    year            = {2016},
}

@article{dolgopolov2024reinforcement,
    author          = {Dolgopolov, Arthur},
    journal         = {Games and Economic Behavior},
    pages           = {84--103},
    title           = {Reinforcement learning in a prisoner's dilemma},
    volume          = {144},
    year            = {2024},
}

@article{dorn2005trading,
    author          = {Dorn, Daniel and Sengmueller, Paul},
    journal         = {Management Science},
    number          = {4},
    pages           = {591--603},
    title           = {Trading as entertainment?},
    volume          = {55},
    year            = {2009},
}

@article{drehmann2013funding,
    author          = {Drehmann, Mathias and Nikolaou, Kleopatra},
    doi             = {10.2139/ssrn.1338092},
    journal         = {Journal of Banking \& Finance},
    number          = {7},
    pages           = {2173--2182},
    publisher       = {Elsevier},
    title           = {Funding liquidity risk: definition and measurement},
    volume          = {37},
    year            = {2013},
}

@article{ElGamal1995,
    author          = {El-Gamal, Mohamed A and Grether, David M},
    doi             = {10.2307/2291506},
    journal         = {Journal of the American Statistical Association},
    number          = {432},
    pages           = {1137--1145},
    title           = {Are people Bayesian? Uncovering behavioral strategies},
    volume          = {90},
    year            = {1995},
}

@article{embrey2017cooperation,
    author          = {Embrey, Matthew and Fr{\'e}chette, Guillaume R and Yuksel, Seda},
    journal         = {American Economic Review},
    number          = {12},
    pages           = {3267--3295},
    title           = {Cooperation in the finitely repeated prisoner’s dilemma},
    volume          = {107},
    year            = {2017},
}

@article{engel2013coevolution,
    author          = {Engel, Christoph and Kurschilgen, Michael},
    doi             = {10.1093/aler/aht010},
    journal         = {American law and economics review},
    number          = {2},
    pages           = {578--609},
    publisher       = {American Law and Economics Association},
    title           = {The coevolution of behavior and normative expectations: An experiment},
    volume          = {15},
    year            = {2013},
}

@article{engel2014first,
    author          = {Engel, Christoph and Beckenkamp, Martin and Gl{\"o}ckner, Andreas and Irlenbusch, Bernd and Hennig-Schmidt, Heike and Kube, Sebastian and Kurschilgen, Michael and Morell, Alexander and Nicklisch, Andreas and Normann, Hans-Theo and others},
    journal         = {International Review of Law and Economics},
    pages           = {126--136},
    publisher       = {Elsevier},
    title           = {First impressions are more important than early intervention: qualifying broken windows theory in the lab},
    volume          = {37},
    year            = {2014},
    doi             = {10.1016/j.irle.2013.07.005},
}

@techreport{engel2014give,
    author          = {Engel, Christoph and Rockenbach, Bettina},
    doi             = {10.2139/ssrn.2519479},
    institution     = {Max Planck Institute for Research on Collective Goods},
    month           = {November},
    number          = {2014/16},
    title           = {Give Everybody a Voice! The Power of Voting in a Public Goods Experiment with Externalities},
    type            = {MPI Collective Goods Preprint},
    url             = {https://ssrn.com/abstract=2519479},
    year            = {2014},
}

@article{engel2015tacit,
    author          = {Engel, Christoph},
    journal         = {Journal of Empirical Legal Studies},
    number          = {3},
    pages           = {537--577},
    publisher       = {Wiley Online Library},
    title           = {Tacit collusion: The neglected experimental evidence},
    volume          = {12},
    year            = {2015},
}

@unpublished{engel2019aim,
    author          = {Engel, Christoph and Kurschilgen, Michael},
    note            = {Draft version, January 2019},
    title           = {Aim High or Aim Low: The Power of Self-Set Normative Goals in a Social Dilemma},
    year            = {2019},
}

@unpublished{Engel2020cart,
    author          = {Engel, Christoph},
    note            = {Working paper},
    title           = {Estimating Heterogeneous Reactions to Experimental Treatments},
    year            = {2020},
}

@article{engel2020fragility,
    author          = {Engel, Christoph and Kurschilgen, Michael},
    doi             = {10.1016/j.joep.2020.102293},
    journal         = {Journal of Economic Psychology},
    pages           = {102293},
    publisher       = {Elsevier},
    title           = {The fragility of a nudge: the power of self-set norms to contain a social dilemma},
    volume          = {81},
    year            = {2020},
}

@article{engel2021managing,
    author          = {Engel, Christoph and Kube, Sebastian and Kurschilgen, Michael},
    doi             = {10.1016/j.jebo.2021.04.029},
    journal         = {Journal of Economic Behavior \& Organization},
    pages           = {111--136},
    publisher       = {Elsevier},
    title           = {Managing expectations: How selective information affects cooperation and punishment in social dilemma games},
    volume          = {187},
    year            = {2021},
}

@article{engel2024makes,
    author          = {Engel, Christoph and Rockenbach, Bettina},
    doi             = {10.1016/j.joep.2024.102712},
    journal         = {Journal of Economic Psychology},
    pages           = {102712},
    publisher       = {Elsevier},
    title           = {What Makes Cooperation Precarious?},
    volume          = {101},
    year            = {2024},
}

@article{engel2024precarious,
    author          = {Engel, Christoph},
    journal         = {Frontiers in Behavioral Economics},
    note            = {Forthcoming},
    title           = {What makes cooperation precarious?},
    year            = {2024},
    pages           = {102712},
    volume          = {101},
    doi             = {10.1016/j.joep.2024.102712},
}

@article{engelrockenbach2020,
    author          = {Engel, Christoph and Rockenbach, Bettina},
    journal         = {Preprint},
    note            = {Max Planck Institute Discussion Paper},
    title           = {Building and milking the public good: Farsightedness in voluntary contribution experiments},
    year            = {2020},
}

@article{ErevRoth1998,
    author          = {Erev, Ido and Roth, Alvin E},
    journal         = {American Economic Review},
    number          = {4},
    pages           = {848--881},
    title           = {Predicting how people play games: Reinforcement learning in experimental games with unique, mixed strategy equilibria},
    volume          = {88},
    year            = {1998},
}

@article{Fallucchi2019,
    author          = {Fallucchi, Francesco and Luccasen, Richard A and Turocy, Theodore L},
    doi             = {10.1007/s40881-018-0060-7},
    journal         = {Journal of the Economic Science Association},
    number          = {2},
    pages           = {238--254},
    title           = {Identifying discrete behavioural types: a re-analysis of public goods game contributions by hierarchical clustering},
    volume          = {5},
    year            = {2019},
}

@article{fallucchi2021identifying,
    author          = {Fallucchi, Francesco and Mercatanti, Andrea and Niederreiter, Jan},
    doi             = {10.1007/s00182-020-00738-w},
    journal         = {International Journal of Game Theory},
    number          = {1},
    pages           = {39--61},
    publisher       = {Springer},
    title           = {Identifying types in contest experiments},
    volume          = {50},
    year            = {2021},
}

@article{Fehr2002,
    author          = {E. Fehr and S. Gächter},
    doi             = {10.1038/415137a},
    journal         = {Nature},
    number          = {6868},
    pages           = {137--140},
    title           = {Altruistic punishment in humans},
    volume          = {415},
    year            = {2002},
}

@article{FehrSchmidt1999,
    author          = {Fehr, Ernst and Schmidt, Klaus M},
    journal         = {The quarterly journal of economics},
    number          = {3},
    pages           = {817--868},
    publisher       = {MIT press},
    title           = {A theory of fairness, competition, and cooperation},
    volume          = {114},
    year            = {1999},
}

@article{Ferraro2010,
    author          = {P. J. Ferraro and C. A. Vossler},
    doi             = {10.2202/1935-1682.2006},
    journal         = {B.E. Journal of Economic Analysis and Policy},
    number          = {1},
    pages           = {53},
    title           = {The source and significance of confusion in public goods experiments},
    volume          = {10},
    year            = {2010},
}

@article{fischbacher2001conditionally,
    author          = {Fischbacher, Urs and G{\"a}chter, Simon and Fehr, Ernst},
    doi             = {10.1016/S0165-1765(01)00394-9},
    journal         = {Economics Letters},
    number          = {3},
    pages           = {397--404},
    publisher       = {Elsevier},
    title           = {Are people conditionally cooperative? Evidence from a public goods experiment},
    volume          = {71},
    year            = {2001},
}

@article{fischbacher2010social,
    author          = {Fischbacher, Urs and G{\"a}chter, Simon},
    doi             = {10.1257/aer.100.1.541},
    journal         = {American Economic Review},
    number          = {1},
    pages           = {541--556},
    publisher       = {American Economic Association},
    title           = {Social preferences, beliefs, and the dynamics of free riding in public good experiments},
    volume          = {100},
    year            = {2010},
}

@article{fischbacher2012behavioral,
    author          = {Fischbacher, Urs and G{\"a}chter, Simon and Quercia, Simone},
    doi             = {10.1016/j.joep.2012.04.002},
    journal         = {Journal of Economic Psychology},
    number          = {4},
    pages           = {897--913},
    publisher       = {Elsevier},
    title           = {The behavioral validity of the strategy method in public good experiments},
    volume          = {33},
    year            = {2012},
}

@article{fischbacher2014heterogeneous,
    author          = {Fischbacher, Urs and Schudy, Simon and Teyssier, Sabrina},
    doi             = {10.1007/s00355-013-0763-x},
    journal         = {Social Choice and Welfare},
    number          = {1},
    pages           = {195--217},
    publisher       = {Springer},
    title           = {Heterogeneous reactions to heterogeneity in returns from public goods},
    volume          = {43},
    year            = {2014},
}

@article{fosgaard2014understanding,
    author          = {Fosgaard, Toke R and Hansen, Lars G and Wengstr{\"o}m, Erik},
    doi             = {10.1016/j.jpubeco.2014.09.004},
    journal         = {Journal of Public Economics},
    pages           = {134--143},
    publisher       = {Elsevier},
    title           = {Understanding the nature of cooperation variability},
    volume          = {120},
    year            = {2014},
}

@article{friedman1971supergames,
    author          = {Friedman, James W.},
    journal         = {The Review of Economic Studies},
    number          = {1},
    pages           = {1--12},
    publisher       = {Oxford University Press},
    title           = {A non-cooperative equilibrium for supergames},
    volume          = {38},
    year            = {1971},
    doi             = {10.1017/cbo9780511528231.011},
}

@article{gaechter2005social,
    author          = {G{\"a}chter, Simon and Th{\"o}ni, Christian},
    journal         = {Journal of the European Economic Association},
    number          = {2-3},
    pages           = {303--314},
    publisher       = {Oxford University Press},
    title           = {Social learning and voluntary cooperation among like-minded people},
    volume          = {3},
    year            = {2005},
    doi             = {10.2139/ssrn.632964},
}

@article{gaechter2007conditional,
    author          = {G{\"a}chter, Simon},
    journal         = {In Economics and Psychology: A Promising New Cross-Disciplinary Field},
    pages           = {19--50},
    publisher       = {MIT Press},
    title           = {Conditional cooperation: Behavioral regularities from the lab and the field and their policy implications},
    year            = {2007},
    doi             = {10.7551/mitpress/2604.003.0006},
}

@article{gaechter2017reciprocity,
    author          = {G{\"a}chter, Simon and K{\"o}lle, Felix and Quercia, Simone},
    doi             = {10.1038/s41562-017-0191-5},
    journal         = {Nature Human Behaviour},
    number          = {9},
    pages           = {650--656},
    publisher       = {Nature Publishing Group},
    title           = {Reciprocity and the tragedies of maintaining and providing the commons},
    volume          = {1},
    year            = {2017},
}

@book{gorton2012misunderstanding,
    author          = {Gorton, Gary},
    publisher       = {Oxford University Press},
    title           = {Misunderstanding Financial Crises: Why We Don't See Them Coming},
    year            = {2012},
}

@article{hendershott2011does,
    author          = {Hendershott, Terrence and Jones, Charles M and Menkveld, Albert J},
    journal         = {The Journal of finance},
    number          = {1},
    pages           = {1--33},
    publisher       = {Wiley Online Library},
    title           = {Does algorithmic trading improve liquidity?},
    volume          = {66},
    year            = {2011},
}

@article{hennig2015handbook,
    author          = {Hennig, Christian},
    journal         = {Pattern Recognition Letters},
    pages           = {53--62},
    publisher       = {Elsevier},
    title           = {What are the true clusters?},
    volume          = {64},
    year            = {2015},
    doi             = {10.1016/j.patrec.2015.04.009},
}

@article{herrmann2009measuring,
    author          = {Herrmann, Benedikt and Th{\"o}ni, Christian},
    doi             = {10.1007/s10683-008-9197-1},
    journal         = {Experimental Economics},
    number          = {1},
    pages           = {87--92},
    publisher       = {Springer},
    title           = {Measuring conditional cooperation: A replication study in Russia},
    volume          = {12},
    year            = {2009},
}

@article{HoCamererChong2007,
    author          = {Ho, Teck-Hua and Camerer, Colin and Chong, Juin-Kuan},
    journal         = {Econometrica},
    number          = {5},
    pages           = {1375--1416},
    title           = {Self-tuning experience-weighted attraction learning in games},
    volume          = {75},
    year            = {2007},
}

@article{Houser2002,
    author          = {D. Houser and R. Kurzban},
    doi             = {10.1257/00028280260344605},
    journal         = {American Economic Review},
    number          = {4},
    pages           = {1062--1069},
    title           = {Revisiting kindness and confusion in public goods experiments},
    volume          = {92},
    year            = {2002},
}

@article{Houser2004,
    author          = {Houser, Daniel and Keane, Michael and McCabe, Kevin},
    doi             = {10.1111/j.1468-0262.2004.00512.x},
    journal         = {Econometrica},
    number          = {3},
    pages           = {781--822},
    title           = {Behavior in a dynamic decision problem: an analysis of experimental evidence using a Bayesian type classification algorithm},
    volume          = {72},
    year            = {2004},
}

@article{isaac1994group,
    author          = {Isaac, R. Mark and Walker, James M.},
    journal         = {The Quarterly Journal of Economics},
    number          = {1},
    pages           = {179--199},
    publisher       = {MIT Press},
    title           = {Group size effects in public goods provision: The voluntary contributions mechanism},
    volume          = {103},
    year            = {1988},
}

@article{iyer2012understanding,
    author          = {Iyer, Rajkamal and Puri, Manju},
    journal         = {American Economic Review},
    number          = {4},
    pages           = {1414--1445},
    title           = {Understanding Bank Runs: The Importance of Depositor-Bank Relationships and Networks},
    volume          = {102},
    year            = {2012},
}

@article{jain1999data,
    author          = {Jain, Anil K. and Murty, M. Narasimha and Flynn, Patrick J.},
    journal         = {ACM Computing Surveys (CSUR)},
    number          = {3},
    pages           = {264--323},
    publisher       = {ACM},
    title           = {Data clustering: A review},
    volume          = {31},
    year            = {1999},
}

@article{Kalweit2020,
    author          = {Kalweit, Gabriel and Huegle, Maria and Werling, Moritz and Boedecker, Joschka},
    journal         = {Advances in Neural Information Processing Systems},
    pages           = {14291--14302},
    title           = {Deep inverse Q-learning with constraints},
    volume          = {33},
    year            = {2020},
}

@article{kamei2012locality,
    author          = {Kamei, Kenju},
    doi             = {10.1016/j.socec.2011.12.005},
    journal         = {Journal of Socio-Economics},
    number          = {2},
    pages           = {207--210},
    publisher       = {Elsevier},
    title           = {From locality to continent: A comment on the generalization of an experimental study},
    volume          = {41},
    year            = {2012},
}

@article{karandikar1998aspirations,
    author          = {Karandikar, Rajiv and Mookherjee, Dilip and Ray, Debraj and Vega-Redondo, Fernando},
    doi             = {10.1006/jeth.1997.2379},
    journal         = {Journal of Economic Theory},
    number          = {2},
    pages           = {292--331},
    publisher       = {Elsevier},
    title           = {Evolving aspirations and cooperation},
    volume          = {80},
    year            = {1998},
}

@article{keser2000conditional,
    author          = {Keser, Claudia and van Winden, Frans},
    journal         = {Scandinavian Journal of Economics},
    number          = {1},
    pages           = {23--39},
    publisher       = {Wiley},
    title           = {Conditional cooperation and voluntary contributions to public goods},
    volume          = {102},
    year            = {2000},
}

@techreport{king2022,
    address         = {Washington, DC},
    author          = {King, Thomas B. and Nesmith, Travis D. and Paulson, Anna and Prono, Todd},
    institution     = {Board of Governors of the Federal Reserve System},
    number          = {2020-009r1},
    title           = {Central Clearing and Systemic Liquidity Risk},
    type            = {Finance and Economics Discussion Series},
    year            = {2022},
}

@article{kirilenko2017flash,
    author          = {Kirilenko, Andrei A. and Kyle, Albert S. and Samadi, Mehrdad and Tuzun, Tugkan},
    journal         = {Journal of Finance},
    number          = {3},
    pages           = {967--998},
    title           = {The Flash Crash: The impact of high-frequency trading on an electronic market},
    volume          = {72},
    year            = {2017},
}

@article{kleiman2018non,
    author          = {Kleiman-Weiner, Max and Tenenbaum, Joshua B. and Zhou, Pang},
    doi             = {10.1111/ectj.12112},
    journal         = {Econometrics Journal},
    number          = {3},
    pages           = {298--315},
    title           = {Non-parametric Bayesian inference of strategies in repeated games},
    volume          = {21},
    year            = {2018},
}

@article{kocher2008conditional,
    author          = {Kocher, Martin G and Cherry, Todd L and Kroll, Stephan and Netzer, Robert J and Sutter, Matthias},
    doi             = {10.1016/j.econlet.2008.07.015},
    journal         = {Economics Letters},
    number          = {3},
    pages           = {175--178},
    publisher       = {Elsevier},
    title           = {Conditional cooperation on three continents},
    volume          = {101},
    year            = {2008},
}

@article{Kosfeld2009,
    author          = {Kosfeld, Michael and Okada, Akira and Riedl, Arno},
    doi             = {10.26481/umamet.2006029},
    journal         = {American Economic Review},
    number          = {4},
    pages           = {1335--55},
    title           = {Institution Formation in Public Goods Games},
    volume          = {99},
    year            = {2009},
}

@article{Kreps1982,
    author          = {Kreps, David M and Milgrom, Paul and Roberts, John and Wilson, Robert},
    journal         = {Journal of Economic Theory},
    number          = {2},
    pages           = {245--252},
    publisher       = {Elsevier},
    title           = {Rational cooperation in the finitely repeated prisoners' dilemma},
    volume          = {27},
    year            = {1982},
}

@article{kumar2009gamblers,
    author          = {Kumar, Alok},
    doi             = {10.1111/j.1540-6261.2009.01483.x},
    journal         = {Journal of Finance},
    number          = {4},
    pages           = {1889--1933},
    title           = {Who gambles in the stock market?},
    volume          = {64},
    year            = {2009},
}

@inproceedings{Kumar2023,
    author          = {Kumar, Sateesh and Zamora, Jonathan and Hansen, Nicklas and Jangir, Rishabh and Wang, Xiaolong},
    booktitle       = {Conference on Robot Learning},
    pages           = {55--66},
    publisher       = {PMLR},
    title           = {Graph inverse reinforcement learning from diverse videos},
    year            = {2023},
}

@article{kurzban2001individual,
    author          = {Kurzban, Robert and Houser, Daniel},
    doi             = {10.1002/per.420},
    journal         = {European Journal of Personality},
    number          = {S1},
    pages           = {S37--S52},
    publisher       = {Wiley Online Library},
    title           = {Individual differences in cooperation in a circular public goods game},
    volume          = {15},
    year            = {2001},
}

@article{kurzban2005experiments,
    author          = {Kurzban, Robert and Houser, Daniel},
    doi             = {10.1073/pnas.0408759102},
    journal         = {Proceedings of the National Academy of Sciences},
    number          = {5},
    pages           = {1803--1807},
    publisher       = {National Academy of Sciences},
    title           = {Experiments investigating cooperative types in humans: A complement to evolutionary theory and simulations},
    volume          = {102},
    year            = {2005},
}

@inproceedings{Kwon2020,
    author          = {Kwon, Minhae and Daptardar, Saurabh and Schrater, Paul R. and Pitkow, Xaq},
    booktitle       = {Advances in Neural Information Processing Systems},
    pages           = {7898--7909},
    title           = {Inverse rational control with partially observable continuous nonlinear dynamics},
    volume          = {33},
    year            = {2020},
}

@incollection{Ledyard1995,
    author          = {J. Ledyard},
    booktitle       = {Handbook of Experimental Economics},
    editor          = {J. Kagel and A. Roth},
    pages           = {253--279},
    publisher       = {Princeton University Press},
    title           = {Public goods: A survey of experimental research},
    year            = {1995},
}

@incollection{ledyard1995handbook,
    author          = {Ledyard, John O.},
    booktitle       = {Handbook of Experimental Economics},
    editor          = {Kagel, John H. and Roth, Alvin E.},
    pages           = {111--194},
    publisher       = {Princeton University Press},
    title           = {Public goods: A survey of experimental research},
    year            = {1995},
}

@article{macy2002learning,
    author          = {Macy, Michael W. and Flache, Andreas},
    doi             = {10.1007/springerreference_301922},
    journal         = {Proceedings of the National Academy of Sciences},
    number          = {Suppl 3},
    pages           = {7229--7236},
    title           = {Learning dynamics in social dilemmas},
    volume          = {99},
    year            = {2002},
}

@article{makowsky2014playing,
    author          = {Makowsky, Michael D and Orman, William H and Peart, Sandra J},
    doi             = {10.1016/j.socec.2014.08.003},
    journal         = {Journal of Behavioral and Experimental Economics},
    pages           = {44--55},
    publisher       = {Elsevier},
    title           = {Playing with other people's money: Contributions to public goods by trustees},
    volume          = {53},
    year            = {2014},
}

@article{montero2022inferring,
    author          = {Montero-Porras, Eladio and Gruji{\'c}, Jelena and Fern{\'a}ndez Domingos, Elias and Lenaerts, Tom},
    journal         = {Scientific reports},
    number          = {1},
    pages           = {7589},
    publisher       = {Nature Publishing Group UK London},
    title           = {Inferring strategies from observations in long iterated prisoner’s dilemma experiments},
    volume          = {12},
    year            = {2022},
    doi             = {10.1038/s41598-022-11654-2},
}

@article{mullainathan2017machine,
    author          = {Mullainathan, Sendhil and Spiess, Jann},
    doi             = {10.1257/jep.31.2.87},
    journal         = {Journal of Economic Perspectives},
    number          = {2},
    pages           = {87--106},
    publisher       = {American Economic Association 2014 Broadway, Suite 305, Nashville, TN 37203-2418},
    title           = {Machine learning: an applied econometric approach},
    volume          = {31},
    year            = {2017},
}

@article{muller2008strategic,
    author          = {Muller, Laurent and Sefton, Martin and Steinberg, Richard and Vesterlund, Lise},
    journal         = {Journal of Economic Behavior \& Organization},
    number          = {3},
    pages           = {782--793},
    publisher       = {Elsevier},
    title           = {Strategic behavior and learning in repeated voluntary contribution experiments},
    volume          = {67},
    year            = {2008},
}

@article{Nasernejad2023,
    author          = {Nasernejad, Payam and Sayed, Tarek and Alsaleh, Rushdi},
    doi             = {10.1080/23249935.2022.2061081},
    journal         = {Transportmetrica A: Transport Science},
    number          = {3},
    pages           = {2061081},
    title           = {Multiagent modeling of pedestrian-vehicle conflicts using adversarial inverse reinforcement learning},
    volume          = {19},
    year            = {2023},
}

@inproceedings{Ng2000,
    author          = {Ng, Andrew Y. and Russell, Stuart J.},
    booktitle       = {International Conference on Machine Learning (ICML)},
    pages           = {2},
    title           = {Algorithms for inverse reinforcement learning},
    volume          = {1},
    year            = {2000},
}

@article{Nikiforakis2008,
    author          = {Nikiforakis, Nikos and Normann, Hans-Theo},
    doi             = {10.2139/ssrn.747144},
    journal         = {Experimental Economics},
    number          = {4},
    pages           = {358--369},
    publisher       = {Springer},
    title           = {A Comparative Statics Analysis of Punishment in Public-Good Experiments},
    volume          = {11},
    year            = {2008},
}

@article{Norenzayan2008,
    author          = {A. Norenzayan and A. F. Shariff},
    doi             = {10.1126/science.1158757},
    journal         = {Science},
    number          = {5898},
    pages           = {58--62},
    title           = {The origin and evolution of religious prosociality},
    volume          = {322},
    year            = {2008},
}

@article{nowak1993pavlov,
    author          = {Nowak, Martin and Sigmund, Karl},
    journal         = {Nature},
    number          = {6432},
    pages           = {56--58},
    publisher       = {Nature Publishing Group},
    title           = {A strategy of win--stay, lose--shift that outperforms tit-for-tat in the Prisoner’s Dilemma},
    volume          = {364},
    year            = {1993},
}

@article{petitjean2011global,
    author          = {Petitjean, Fran{\c{c}}ois and Ketterlin, Alain and Gan{\c{c}}arski, Pierre},
    doi             = {10.1016/j.patcog.2010.09.013},
    journal         = {Pattern recognition},
    number          = {3},
    pages           = {678--693},
    publisher       = {Elsevier},
    title           = {A global averaging method for dynamic time warping, with applications to clustering},
    volume          = {44},
    year            = {2011},
}

@article{Rabin1993,
    author          = {Rabin, Matthew},
    doi             = {10.2307/j.ctvcm4j8j.15},
    journal         = {The American economic review},
    pages           = {1281--1302},
    publisher       = {JSTOR},
    title           = {Incorporating fairness into game theory and economics},
    year            = {1993},
}

@article{RangelCamererMontague2008,
    author          = {Rangel, Antonio and Camerer, Colin and Montague, P. Read},
    doi             = {10.1038/nrn2357},
    journal         = {Nature Reviews Neuroscience},
    number          = {7},
    pages           = {545--556},
    title           = {A framework for studying the neurobiology of value-based decision making},
    volume          = {9},
    year            = {2008},
}

@article{Rosenberg2021,
    author          = {Rosenberg, Matthew and Zhang, Tony and Perona, Pietro and Meister, Markus},
    doi             = {10.7554/eLife.66175},
    journal         = {eLife},
    pages           = {e66175},
    title           = {Mice in a labyrinth show rapid learning, sudden insight, and efficient exploration},
    volume          = {10},
    year            = {2021},
}

@article{rousseeuw1987silhouettes,
    author          = {Rousseeuw, Peter J.},
    doi             = {10.1016/0377-0427(87)90125-7},
    journal         = {Journal of Computational and Applied Mathematics},
    pages           = {53--65},
    publisher       = {Elsevier},
    title           = {Silhouettes: a graphical aid to the interpretation and validation of cluster analysis},
    volume          = {20},
    year            = {1987},
}

@article{su2016identifying,
    author          = {Su, Liangjun and Shi, Zhentao and Phillips, Peter CB},
    doi             = {10.2139/ssrn.2448189},
    journal         = {Econometrica},
    number          = {6},
    pages           = {2215--2264},
    publisher       = {Wiley Online Library},
    title           = {Identifying latent structures in panel data},
    volume          = {84},
    year            = {2016},
}

@article{thoni2012microfoundations,
    author          = {Th{\"o}ni, Christian and Tyran, Jean-Robert and Wengstr{\"o}m, Erik},
    doi             = {10.1016/j.jpubeco.2012.04.003},
    journal         = {Journal of Public Economics},
    number          = {7--8},
    pages           = {635--643},
    publisher       = {Elsevier},
    title           = {Microfoundations of social capital},
    volume          = {96},
    year            = {2012},
}

@article{thoni2018conditional,
    author          = {Th{\"o}ni, Christian and Volk, Stefan},
    doi             = {10.1016/j.econlet.2018.06.022},
    journal         = {Economics Letters},
    pages           = {37--40},
    publisher       = {Elsevier},
    title           = {Conditional cooperation: Review and refinement},
    volume          = {171},
    year            = {2018},
}

@article{trautmann2011bank,
    author          = {Trautmann, Stefan T. and Vlahu, Razvan},
    journal         = {Review of Finance},
    number          = {2},
    pages           = {971--1000},
    title           = {Bank-Run Psychology},
    volume          = {17},
    year            = {2011},
}

@article{van2014implementing,
    author          = {van Miltenburg, Niels and Buskens, Vincent and Barrera, Davide and Raub, Werner},
    doi             = {10.18352/ijc.397},
    journal         = {International Journal of the Commons},
    number          = {1},
    pages           = {47--78},
    title           = {Implementing punishment and reward in the public goods game: The effect of individual and collective decision rules},
    volume          = {8},
    year            = {2014},
}

@article{Virtanen2020SciPy,
    author          = {Virtanen, Pauli and Gommers, Ralf and Oliphant, Travis E. and Haberland, Matt and Reddy, Tyler and Cournapeau, David and Burovski, Evgeni and Peterson, Pearu and Weckesser, Warren and Bright, Jonathan and {van der Walt}, Stéfan J. and Brett, Matthew and Wilson, Joshua and Millman, K. Jarrod and Mayorov, Nikolay and Nelson, Andrew R. J. and Jones, Eric and Kern, Robert and Larson, Eric and Carey, Christopher J. and Polat, {\.{I}}lhan and Feng, Yu and Moore, Eric W. and VanderPlas, Jake and Laxalde, Denis and Perktold, Josef and Cimrman, Robert and Henriksen, Ian and Quintero, E. A. and Harris, Charles R. and Archibald, Anne M. and Ribeiro, Ant\^{o}nio H. and Pedregosa, Fabian and {van Mulbregt}, Paul and {SciPy 1.0 Contributors}},
    doi             = {10.1038/s41592-019-0686-2},
    journal         = {Nature Methods},
    number          = {3},
    pages           = {261--272},
    title           = {{SciPy} 1.0: Fundamental Algorithms for Scientific Computing in Python},
    volume          = {17},
    year            = {2020},
}

@article{volk2012temporal,
    author          = {Volk, Stephan and Th{\"o}ni, Christian and Ruigrok, Winfried},
    doi             = {10.1016/j.jebo.2011.10.006},
    journal         = {Journal of Economic Behavior \& Organization},
    number          = {2},
    pages           = {664--676},
    publisher       = {Elsevier},
    title           = {Temporal stability and psychological foundations of cooperation preferences},
    volume          = {81},
    year            = {2012},
}

@inproceedings{von2012clustering,
    author          = {Von Luxburg, Ulrike and Williamson, Robert C and Guyon, Isabelle},
    booktitle       = {Proceedings of ICML workshop on unsupervised and transfer learning},
    organization    = {JMLR Workshop and Conference Proceedings},
    pages           = {65--79},
    title           = {Clustering: Science or art?},
    year            = {2012},
}

@article{wang2024confusion,
    author          = {Wang, Guangrong and Li, Jianbiao and Wang, Wenhua and Niu, Xiaofei and Wang, Yue},
    journal         = {Proceedings of the National Academy of Sciences},
    number          = {10},
    pages           = {e2310109121},
    publisher       = {National Acad Sciences},
    title           = {Confusion cannot explain cooperative behavior in public goods games},
    volume          = {121},
    year            = {2024},
}

@article{weber2018dispositional,
    author          = {Weber, Timo O and Weisel, Ori and G{\"a}chter, Simon},
    doi             = {10.1038/s41467-018-04775-8},
    journal         = {Nature Communications},
    number          = {1},
    pages           = {2390},
    publisher       = {Nature Publishing Group},
    title           = {Dispositional free riders do not free ride on punishment},
    volume          = {9},
    year            = {2018},
}

@article{wermers2000mutual,
    author          = {Wermers, Russ},
    doi             = {10.2469/dig.v31.n1.836},
    journal         = {Journal of Finance},
    number          = {4},
    pages           = {1655--1695},
    title           = {Mutual fund performance: An empirical decomposition into stock-picking talent, style, transactions costs, and expenses},
    volume          = {55},
    year            = {2000},
}

@article{Yamaguchi2018,
    author          = {Yamaguchi, Shoichiro and Naoki, Honda and Ikeda, Muneki and Tsukada, Yuki and Nakano, Shunji and Mori, Ikue and Ishii, Shin},
    doi             = {10.1101/129007},
    journal         = {PLoS Computational Biology},
    number          = {5},
    pages           = {e1006122},
    title           = {Identification of animal behavioral strategies by inverse reinforcement learning},
    volume          = {14},
    year            = {2018},
}

@article{Zelmer2003,
    author          = {Zelmer, Jennifer},
    doi             = {10.1023/a:1026277420119},
    journal         = {Experimental Economics},
    number          = {3},
    pages           = {299--310},
    publisher       = {Springer},
    title           = {Linear Public Goods Experiments: A Meta-Analysis},
    volume          = {6},
    year            = {2003},
}

@article{zheng2024evolution,
    author          = {Zheng, Guozhong and Zhang, Jiqiang and Deng, Shengfeng and Cai, Weiran and Chen, Li},
    doi             = {10.1016/j.chaos.2023.115568},
    journal         = {Chaos, Solitons and Fractals},
    pages           = {115568},
    publisher       = {Elsevier},
    title           = {Evolution of cooperation in the public goods game with Q-learning},
    volume          = {188},
    year            = {2024},
}

@article{Zhu2024,
    author          = {Zhu, Hao and De La Crompe, Brice and Kalweit, Gabriel and Schneider, Artur and Kalweit, Maria and Diester, Ilka and Boedecker, Joschka},
    journal         = {Transactions on Machine Learning Research},
    note            = {Available online: \url{https://openreview.net/forum?id=hrKHkmLUFk}},
    title           = {Multi-intention Inverse Q-learning for Interpretable Behavior Representation},
    year            = {2024},
}

@article{zhu2025fitting,
    author          = {Zhu, Hao and Hoffmann, Jasper and Zhang, Baohe and Boedecker, Joschka},
    journal         = {arXiv preprint arXiv:2511.04454v1},
    pages           = {353-376},
    title           = {Fitting Reinforcement Learning Model to Behavioral Data under Bandits},
    year            = {2025},
}
}

\clearpage
\begin{appendices}
\onecolumn
\setcounter{figure}{0}   
\setcounter{table}{0} 
\renewcommand{\thesection}{\Alph{section}}%
\renewcommand{\thesubsection}{\thesection.\arabic{subsection}}
\renewcommand\thefigure{S.\arabic{figure}} 
\renewcommand\thetable{S.\arabic{table}}
\begin{center}
    \begin{minipage}{\textwidth}
        \centering
        \fontsize{16}{20}\selectfont
        %\thetitle\\
        Supplementary Materials
    \end{minipage}
\end{center}
\vspace{.5cm}

\section{Details on Methods}
\label{apx:methods}

Several approaches have been proposed to classify behavioral types in PGGs. These approaches differ primarily along two dimensions: their reliance on theoretical assumptions (vs. purely data-driven approaches), and how they accommodate the inherent temporal structure of behavioral data.

At the theory-driven end, {\em finite mixture models} are widely used: Applications include PGGs \citep{bardsley2007experimetrics}, beauty contest experiments \citep{BoschDomenech2010}, lottery-choice tasks \citep{Conte2011}, and private-information games \citep{Brocas2014}. 

Similarly, {\em Bayesian mixture models} have gained traction for identifying latent behavioral types, particularly in PGGs \citep{Houser2004} and probabilistic updating tasks \citep{ElGamal1995}. These methods typically follow a three-stage procedure: first, specifying a parametric model of individual decision-making; second, estimating individual-specific parameters via Bayesian inference; and third, clustering these parameters post-estimation to delineate distinct behavioral types.

{\em C-Lasso} \citep{su2016identifying} extends this logic by merging estimation and clustering into a unified penalized regression framework. Instead of sequentially estimating parameters and clustering afterward, C-Lasso shrinks parameter estimates toward common values and thus implicitly groups subjects during estimation. This method has notably identified heterogeneous behaviors in contest experiments \citep{fallucchi2021identifying}.

At the opposite, data-driven end of the spectrum lie unsupervised machine learning methods, such as {\em clustering algorithms} employed by \citet{Fallucchi2019, bolle2021behavioral} in analyzing social dilemmas. From an econometric perspective, these clustering methods function as dimensionality-reduction tools, uncovering latent behavioral patterns without predefined assumptions on distributions or preferences.

Instead of comparing points strictly by index, {\em DTW} aligns key structural moments—such as peaks—by warping the time dimension to minimize the overall alignment cost \citep[see][for details]{Berndt1994}.
Formally, for two time series, \(\mathbf{x} = \{x_1, x_2, \dots, x_T\}\) and \(\mathbf{y} = \{y_1, y_2, \dots, y_T\}\), DTW constructs a cost matrix \(D\) where each element \(D(i, j)\) represents the cumulative cost of aligning \(x_i\) with \(y_j\). The recurrence relation for computing \(D(i, j)\) is given by
\[
\begin{aligned}
D(i, j) &= d(x_i, y_j) \\
        &\quad + \min \left\{ D(i-1, j),\, D(i, j-1),\, D(i-1, j-1) \right\},
\end{aligned}
\]
where the local distance \(d(x_i, y_j)\) is typically defined as the squared difference \((x_i - y_j)^2\). The optimal DTW distance is then given by \(D(T, T)\), found at the bottom-right corner of the matrix, which represents the minimal cumulative alignment cost.

\clearpage
\section{Types in Behavioral Finance}

\begin{table}[!htbp]
    \centering
        \caption{Use of Time-Series Data and Type Identification Methods in Behavioral Finance.}
{\resizebox{\textwidth}{!}{
\begin{tabular}{p{3cm} p{2cm} p{4.2cm} p{4.2cm} p{3.2cm}}
\toprule
\textbf{Paper} & \textbf{Time Series Used} & \textbf{How Time Series Are Processed} & \textbf{Type Identification Method} & \textbf{Types Identified} \\
\midrule
\citet{barber2001boys} & Trading history & Aggregated into turnover, returns, performance measures & Behavioral signature inference (no clustering) & Overconfident vs. less overconfident \\

\citet{barber2001boys}
& Trading history 
& Aggregated into turnover, returns, performance measures 
& Behavioral signature inference (no clustering) 
& Overconfident vs. less overconfident \\

\citet{kumar2009gamblers}
& Trade sequences 
& Collapsed into lottery-preference index 
& Preference-based econometric scoring 
& Lottery-seeking vs. non-gamblers \\

\citet{dorn2005trading}
& Trading frequency 
& Reduced to counts and averages 
& Descriptive segmentation 
& Entertainment traders vs. others \\

\citet{calvet2009financial}
& Portfolio histories 
& Used in the structural model of portfolio choice
& Structural posterior classification 
& Sophisticated vs. unsophisticated investors\\

\citet{kirilenko2017flash}
& HFT order flow 
& Engineered features (inventory, cancellations, message rates) 
& Unsupervised clustering (k-means)
& Market makers, opportunistic traders, directional traders \\

\citet{brogaard2014high}
& HFT quotes \& inventory 
& Aggregated into liquidity-supply ratios 
& Rule-based classification 
& Liquidity providers vs. liquidity takers \\

\citet{brown1996tournaments}
& Fund returns 
& Estimated changing factor loadings and risk exposure over time
& Structural/Regression-based style analysis
& Risk-shifters, conservative, tournament-driven managers \\

\citet{wermers2000mutual}
& Fund returns \& holdings 
& Performance decomposition over time 
& Component-based classification 
& Stock pickers vs. style-driven funds \\

\citet{iyer2012understanding}
& Withdrawal timing 
& Hazard/survival modeling of sequences 
& Regression-based segmentation 
& Early, conditional, late withdrawers \\

\citet{trautmann2011bank}
& Experimental round decisions 
& Strategy-method elicitation \& decision sequences 
& Rule-based or simple mixture modeling 
& Run-prone, cautious, strategic waiters \\

\citet{gorton2012misunderstanding}
& Crisis dynamics 
& Descriptive analysis of crisis time series 
& Conceptual typology 
& Run-prone, strategic waiters, fundamentalists \\
\bottomrule
\end{tabular}}}
\vspace{4pt}
\parbox{\linewidth}{\footnotesize\raggedright
\textit{Note:} This literature typically reduces behavioral trajectories to engineered features, summary statistics, or factor loadings, instead of using the complete time series.}
\label{tab:behavioral_finance}
\end{table}

\clearpage
\section{Implementation Details of the HIQL Model}

\paragraph{Fitting Process:}
We discretize the continuous range $[0,1]$ of own contributions and others' contributions into five bins, which we denote as the action set $  \mathcal{A} \;=\; \{\, a_1, a_2, a_3, a_4, a_5 \} \quad (\text{five bins in }[0,1]).$
Similarly, we denote the state set as $ \mathcal{S} \;=\; \{\, s_1, s_2, s_3, s_4, s_5 \} \quad (\text{five bins in }[0,1]).$
The binning has the primary purpose of reducing the size of the Q-table. The original action set has, on average, 11 bins (in the canonical public good game, participants have an endowment of 10 tokens, and may contribute between 0 and 10 tokens to the joint project). As states are averages, their space is even larger. As defined earlier, the input for the model is time series trajectories (per participant) of state-action-next state triplets.

Following \citet{Zhu2024}, we employ a multi-stage fitting procedure.
For all steps, we set the discount factor to $\gamma = 0.99$. 

In the first stage, we run multiple HIQL fits with different numbers of intentions $K = 2, \ldots, 5$ on the non-clustered dataset to obtain a global fit. 
We split this data into a train/test set. Reward functions are estimated on the training set, and predictive performance is evaluated on the test set using the out-of-sample/test log-likelihood. The choice of $K$ is based on this predictive performance and the Bayesian Information Criterion (BIC). 
For each fitting, we perform 10 different initializations and select the best-performing initialization. 
The initial intention distribution $\Pi$ is initialized uniformly, and the intention transition matrix $\Lambda$ is initialized according to \citet{Zhu2024} as $\Lambda = 0.95 \times I + N(0, 0.05 \times I)$ where $N$ denotes the normal distribution and $I \in \mathbb{R}^{K \times K}$ is the identity matrix. This initial $\Lambda$ is then normalized so that each row adds up to 1.

In the second stage of the fitting procedure, we obtain an independent but aligned HIQL fit for each cluster. 
Specifically, we initialize each cluster’s HIQL model using the best global parameters (the global prior over initial latent intentions and the global transition probabilities between intentions across time) learned from all data combined. 
Using these as initialization ensures that all cluster-specific models start from a consistent baseline of how intentions are expected to emerge and evolve, while allowing each cluster to adapt its own intention-specific policies and rewards to reflect the behavioral characteristics of participants within that cluster.

As in \citet{Zhu2024}, we apply a five-fold cross-validation approach, and additionally implement a repeated stratified fold split. The stratification is based on the average contribution of individuals. We bin the clusters into three average contribution bins and ensure that both the training and test sets are sampled from all bins. 
Furthermore, we repeat this process ten times within each fold to obtain stable posterior estimates per participant.

\paragraph{Assessing Model Fit:}
Model fit is evaluated using the log-likelihood (LL), which quantifies how well the learned model explains the observed actions given the states. Because IRL models behavior probabilistically—estimating the likelihood that a participant selects a given action under a latent policy—the LL serves as a natural and sufficient measure of model performance. 
Higher (less negative) log-likelihood values indicate that the model assigns greater probability to the actions actually observed, meaning it captures participants’ decision patterns more accurately. 
Conversely, lower log-likelihood values reflect greater behavioral noise or variability, as the model must distribute probability more diffusely across possible actions.

\paragraph{Aligning Latents:}
The main challenge in any latent-variable model lies in interpreting and aligning the latent dimensions. Because we estimate separate HIQL models for each behavioral cluster, the latent intentions are not automatically aligned across clusters—or across different folds and repetitions in cross-validation.
To establish consistent labeling of intentions, we first visualize each participant’s input data (states and actions) together with their inferred latent intentions. From these visualizations, we interpret the two dominant latent dimensions as reflecting \emph{cooperative} and \emph{defective} intentions. Accordingly, peaks and valleys in the action time series correspond to rises and falls in the inferred cooperation intention.
Next, we apply a peak-detection algorithm \citep{Virtanen2020SciPy} to identify peaks in both the action and latent-intention time series. Within each cluster, we then quantify how often each posterior intention aligns with the detected behavioral peaks. The intention showing the highest alignment is designated as \emph{intention~1}.

\clearpage
\section{Details on the Data}

\begin{table}[!htbp]
    \centering
\caption{Data Overview by Period Length.}
\label{tab:data_details}

{\begin{tabular}{@{}l@{\quad}lcc@{}}
\hline
Period Length & Study & Group Size & Participants \\ \hline

\multirow{3}{*}{7} 
& \citet{Diederich2016} & 10 & 410 \\
& \citet{Diederich2016} & 40 & 200 \\
& \citet{Diederich2016} & 100 & 500 \\
\cmidrule(l){2-4}
& \textit{Subset Total} & & \textbf{1110} \\ \midrule

\multirow{4}{*}{10} 
& \citet{engel2021managing} & 4 & 236 \\
& \citet{engel2014give} & 4 & 102 \\
& \citet{Nikiforakis2008} & 4 & 24 \\
& \citet{Diederich2016} & 40 & 200 \\ % (included for completeness if needed)
\cmidrule(l){2-4}
& \textit{Subset Total} & & \textbf{562} \\ \midrule

\multirow{3}{*}{20} 
& \citet{engel2024makes} & 3 & 30 \\
& \citet{engel2024makes} & 5 & 252 \\
& \citet{Kosfeld2009} & 4 & 216 \\
\cmidrule(l){2-4}
& \textit{Subset Total} & & \textbf{498} \\ \midrule

\multirow{4}{*}{30} 
& \citet{engel2014first} & 4 & 228 \\
& \citet{engel2013coevolution} & 4 & 20 \\
& \citet{engel2019aim} & 4 & 288 \\
& \citet{engel2020fragility} & 4 & 432 \\
\cmidrule(l){2-4}
& \textit{Subset Total} & & \textbf{968} \\ \midrule

\textbf{Grand Total} & & & \textbf{2938} \\ \hline
\end{tabular}}

\vspace{4pt}
\parbox{\linewidth}{\footnotesize\raggedright
\textit{Note:} Raw data were collected from published public good experiments.
We categorize studies by the number of periods (round lengths).
Inclusion criteria: (1) standard linear public good game; (2) feedback on others’ contributions after each round.}
    
\end{table}

\begin{figure*}[!htbp]
\centering
\includegraphics[width=\linewidth]{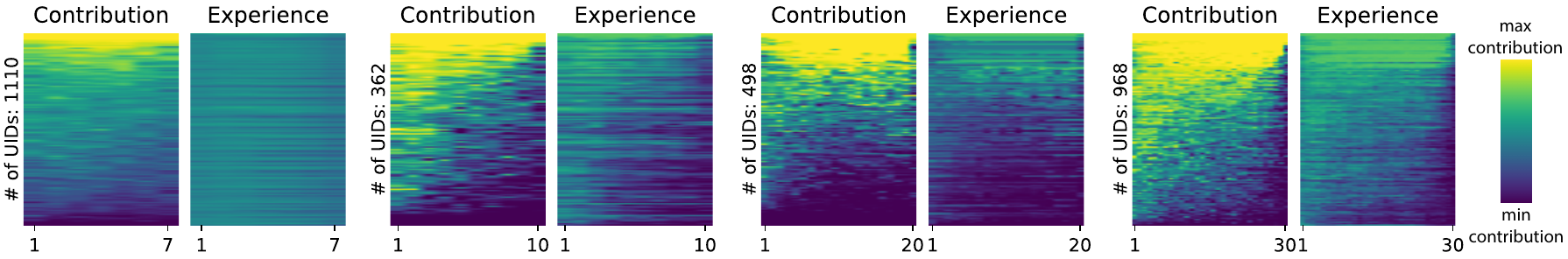}
\caption{{\bf Visualization of the Raw Data.}The y-axis represents participant identifiers (UIDs), while the x-axis tracks the rounds played.
Color intensity indicates the size of contributions, normalized between 0 and 1.
The panels labeled `Contribution' reflect each player's own contribution.
The panels labeled `Experience' represent the average contribution of the other players in the group during the preceding round.
Within subplot-pairs, UIDs are sorted by their average contribution.
Notably, there is heterogeneity with respect to the number of rounds played: The data is divided into subsets with games lasting 7, 10, 20, or 30 rounds.
Furthermore, there exists heterogeneity concerning game parameters.
Most notably, one study played a 7-round game (first column) with an exceptionally large group size of 100, resulting in unusually homogeneous average contributions.}
\label{fig:data}
\end{figure*}

\clearpage
\section{Details on the Clustering Analysis}

\subsection{Finding the Optimal Number of Clusters}
\label{apx:n_clusters}

\begin{figure}[H]
    \centering
    \includegraphics[width=0.35\linewidth]{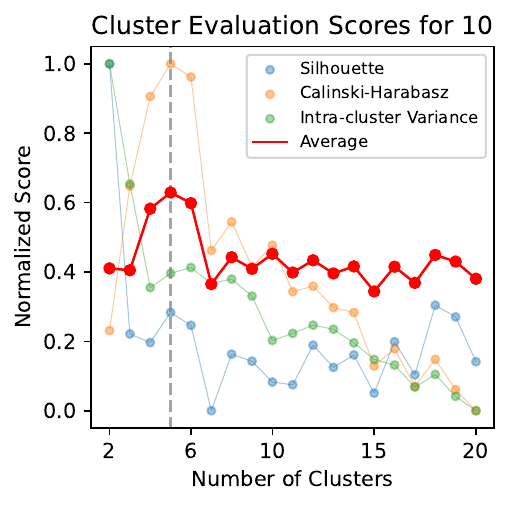}
    \caption{{\bf Determining the Optimal Number of Clusters $k$.} Normalized cluster evaluation scores (Silhouette, Davies–Bouldin, and Intra-cluster Variance) across different numbers of clusters.
The vertical grey line indicates the selected solution with $k = 6$, which provides a trade-off between the evaluation metrics
and maintains meaningful cluster sizes (each containing at least 30 members).}
    \label{fig:optimal-k}
\end{figure}

To identify the appropriate number of behavioral clusters, we use three validity indices compatible with DTW-based clustering: the Silhouette Score \citep{rousseeuw1987silhouettes}, the Calinski–Harabasz Index \citep{calinski1974dendrite}, and Intra-cluster Variance \citep{jain1999data}. Together, they evaluate cluster separation and the internal consistency of behavior.
We compute each metric for $k = 2$ to $k = 20$, min–max normalize the results, and invert Intra-cluster Variance so that higher values consistently indicate better clustering. As shown in Fig.~\ref{fig:optimal-k}, the average normalized score is highest for $k=4$–$6$, with a maximum at $k=5$. Beyond $k=6$, performance becomes unstable, and clusters shrink to sizes that limit interpretation. We therefore adopt $k=6$ as a balanced solution—statistically strong while maintaining clear and interpretable behavioral groups. This approach aligns with recommendations that cluster validity indices serve as guidance rather than strict optimization rules \citep{hennig2015handbook}.

Our dataset includes experiments with heterogeneous designs—different round lengths and group sizes—which shape behavioral dynamics. For example, the 7-round studies involve unusually large groups \citep{isaac1994group}, the 20-round studies permit punishment and related enforcement-responsive strategies, and the 30-round studies resemble long-horizon interactions. By contrast, 10-round public goods games with 4-person groups represent the modal experimental design \citep{ledyard1995handbook, chaudhuri2011sustaining}. To avoid overfitting to design-specific behavior, we use this standard subset to evaluate cluster validity and then apply the resulting six-cluster taxonomy across all studies. This ensures that our behavioral types reflect generalizable strategic patterns rather than experimental idiosyncrasies.

\clearpage

\subsection{Assessing Cluster Stability and Heterogeneity}
\label{apx:cluster_heterogeneity}

We evaluate cluster quality along two dimensions: (1) stability under resampling and (2) within-cluster heterogeneity. 
Stability assesses robustness to data perturbations, while heterogeneity tests whether clustering overfits by forcing all trajectories into six groups without a residual category.

We conduct non-parametric bootstrap resampling (100 replications, 80\% subsample) to assess assignment robustness. 
Table~\ref{tab:cluster_diag_sidebyside} reports subject-level stability, defined as the share of replications in which participants return to their baseline cluster. 
On average, participants return to their dominant cluster in 29.9\% of replications (SD = 3.3 pp), well above the 16.7\% expected under random assignment in a six-cluster design.
Stability is similar across clusters (range: 27.2\%--31.2\%), suggesting that all clusters represent comparably robust behavioral patterns rather than noise.
At the partition level, bootstrap solutions closely match the original clustering (ARI = 0.861, SD = 0.061; NMI = 0.868, SD = 0.047), confirming stable boundaries and assignments.

Despite this stability, substantial heterogeneity remains within clusters.
We quantify heterogeneity using silhouette scores, which range from $-1$ (closer to another cluster) to $+1$ (strong fit to the assigned cluster).
As shown in Table~\ref{tab:cluster_diag_sidebyside2}, cluster coherence varies widely: Clusters 1 and 6 exhibit strong internal structure (mean silhouette $>0.46$), while Clusters 2 and 4 show weak separation (mean silhouette $<0.05$).
Furthermore, 26.9\% of participants have negative silhouette scores, meaning their trajectories are closer (on average) to another cluster than to the one they are assigned. 
Distance-based metrics tell a similar story. The average distance between trajectories within the same cluster is $1.87$ ($SD=0.49$), while the average distance between trajectories in different clusters is only modestly larger at $2.92$ ($SD=1.00$).
Consistent with this overlap, a DTW-based variance decomposition partitions total trajectory dispersion into between- and within-cluster components. Clusters account for 29.5\% of total variance—reflecting differences between cluster-average trajectories—while the remaining 70.5\% reflects variation among individuals within the same cluster.

Thus, clusters summarize dominant behavioral patterns while leaving substantial individual-level variation unexplained. This is consistent with extensive evidence from the PGG literature that individual contribution paths exhibit considerable idiosyncratic variation. The fact that clusters are stable across resampling yet retain substantial within-cluster heterogeneity indicates that they capture systematic tendencies without artificially suppressing behavioral noise.

\begin{table}[!htbp]
\centering
\caption{\textbf{Subject-Level Cluster Stability}}
\label{tab:cluster_diag_sidebyside}

\begin{tabular}{lcccccc|c}
\hline
 & Cl. 1 & Cl. 2 & Cl. 3 & Cl. 4 & Cl. 5 & Cl. 6 & \textbf{Overall} \\
\hline
Mean Stability (\%) & 30.5 & 31.2 & 29.2 & 31.2 & 30.1 & 27.2 & 29.9 \\
SD (pp)             &  3.7 &  2.0 &  3.1 &  3.8 &  3.2 &  2.9 &  3.3 \\
\hline
\end{tabular}

\vspace{4pt}
\footnotesize\raggedright
Note: Mean stability indicates the share of bootstrap replications (100 replications, 80\% subsample)
in which subjects return to their baseline cluster.
SD reflects variability across subjects within each cluster (percentage points, pp).
Random assignment benchmark: 16.7\%.
\end{table}

\begin{table}[!htbp]
\centering
\caption{\textbf{Within-Cluster Heterogeneity and Fit Quality}}
\label{tab:cluster_diag_sidebyside2}

\begin{tabular}{lcccccc|c}
\hline
 & Cl. 1 & Cl. 2 & Cl. 3 & Cl. 4 & Cl. 5 & Cl. 6 & \textbf{Overall} \\
\hline
Mean Silhouette & 0.47 & 0.00 & 0.10 & 0.05 & 0.29 & 0.48 & 0.13 \\
SD              & 0.08 & 0.13 & 0.11 & 0.15 & 0.11 & 0.08 & 0.21 \\
\hline
\end{tabular}

\vspace{4pt}
\footnotesize\raggedright
Note: Silhouette scores range from $-1$ to $+1$; negative values indicate participants closer to an alternative cluster.
Within-cluster distance: $M=1.87$ ($SD=0.49$); between-cluster distance: $M=2.92$ ($SD=1.00$);
separation ratio: $1.56$. Overall, 26.9\% have negative silhouette scores.
\end{table}

\subsection{Effect of Distance Metrics on Clustering Results}
\label{apx:euclidean_vs_dtw}

\begin{figure}[H]
    \centering
    \includegraphics[width=0.9\linewidth]{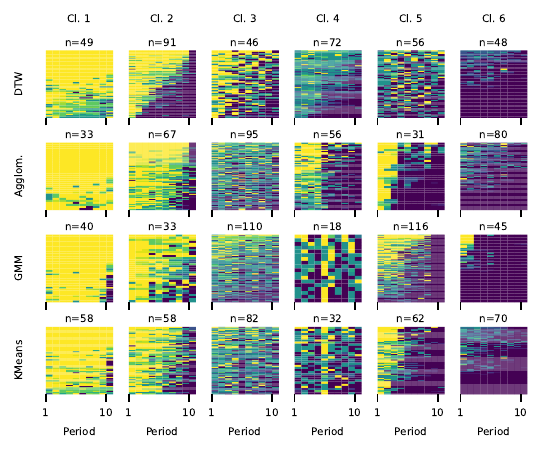}
    \caption{{\bf Comparison of Distance Metrics.} Clustering based on DTW k-means (top row) is contrasted with Euclidean-based clustering (bottom row), including agglomerative clustering,
    Gaussian mixture models and k-means. Euclidean distance enforces strict pointwise alignment across rounds, producing vertically aligned
    contribution patterns. A behavioral profile defined by a sharp drop from high to very low contributions is consistently detected
    across methods but may be assigned to different clusters under Euclidean agglomerative clustering.}
    \label{fig:compare-dist-10}
\end{figure}

We illustrate that the choice of distance metric makes a qualitative difference in the clustering outcomes.
Euclidean distance allows only for local alignment, whereas DTW enables global alignment—meaning that time series exhibiting the same overall pattern but shifted in time can be grouped into the same cluster.
We generate a comprehensive grid of comparisons: DTW paired with k-means, Euclidean distance paired with k-means, Gaussian Mixture Models (GMM), and agglomerative clustering.

Fig.~\ref{fig:compare-dist-10} shows that all Euclidean-based results display vertically aligned contribution patterns.

For illustrative purposes, we focus on the cluster type characterized by a sharp transition from very high to very low contributions. This cluster (Cluster 2 under DTW) is most clearly visible among the Euclidean-based clusters in the agglomerative (row 2) outcomes.
We therefore select these two methods—DTW with k-means and Euclidean with agglomerative clustering—for direct comparison in the main text (Fig.~\ref{fig:distance_metric_comparison}).

\begin{figure}
    \centering
    \includegraphics[width=\linewidth]{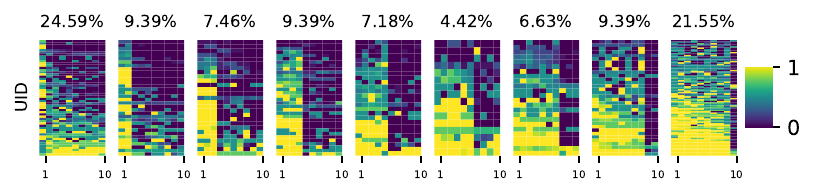}
    \caption{{\bf Distribution of Optimal Threshold Points Across Rounds.} Each subplot represents the percentage of players who showed their most dramatic change in contribution level at that round.
Heatmaps of contributions are shown, subsetted by the round in which the switching process occurs.}
    \label{fig:switching_partitions}
\end{figure}

%\clearpage
%\paragraph{Analysis of Threshold Switching Behavior}
We further analyze the threshold-switching hypothesis. Cluster 2 in the DTW results suggests the existence of a behavioral type that switches from very high to very low contributions quite abruptly, and that this switch can occur at any point in time. 
An alternative hypothesis is that switching occurs only at specific, time-locked points in the game.
To test these hypotheses, we identify, for each participant, the round in which the most dramatic change in contributions occurs. 
We then group participants based on these switching points (Fig.~\ref{fig:switching_partitions}).
Clusters 1 and 9 together contain nearly 50\% of participants and are visually the least distinct, while the remaining clusters each include between 4.4\% and 9.4\% of participants.
Notably, Clusters 1 and 9 do not display the strong yellow–dark blue contrast that characterizes the sharp ``high-to-low" transition observed in Cluster 2 (DTW). While the relative size of these clusters is indeed higher than that of the others, the absence of a clear color transition indicates that they do not capture the same extreme switching behavior.
Across the remaining clusters, each cluster has a nonzero share of participants, suggesting that switching can occur at any round.
Overall, this analysis supports the interpretation that extreme switching behavior is not tied to a specific point in time but can arise idiosyncratically at any stage of the game.

\clearpage

\subsection{Comparison of Clustering Algorithms}
\label{apx:cluster_algorithms}

\begin{figure}[H]
    \centering
    \includegraphics[width=.99\linewidth]{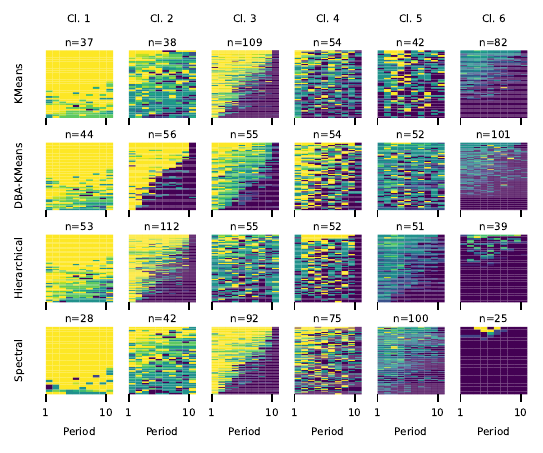}
    \caption{{\bf Comparison of Clustering Algorithms in Combination with DTW.} We examine combinations of DTW with K-means, hierarchical clustering, and spectral clustering.
    For this analysis, the number of clusters is fixed to six, focusing on the data subset with 10 rounds.
    The results indicate that spectral clustering most distinctly separates the data.}
    \label{fig:dtw_comp}
\end{figure}

Having identified DTW as the most suitable distance metric for our data, we evaluated its performance in combination with four clustering algorithms: K-means, DBA K-means, hierarchical clustering, and spectral clustering (see Fig.~\ref{fig:dtw_comp}). Overall, the resulting clusters are highly similar across all algorithms, indicating that the underlying structure of the data is stable and can be reliably identified by different methods when the appropriate distance metric is applied.

Qualitatively, spectral clustering produced the most clearly distinct \textit{freeriding} cluster as well as the most pronounced \textit{unconditional contribution} cluster. Therefore, we selected spectral clustering for all subsequent analyses.

\clearpage
\subsection{Comparison of Partitioning Methods from the Literature}

\begin{figure}[H]
    \centering
    {%
\begin{minipage}{.8\linewidth}
\centering
\fbox{\makebox[4.3cm][c]{\textbf{A.} DTW}}\\
\includegraphics[width=\linewidth]{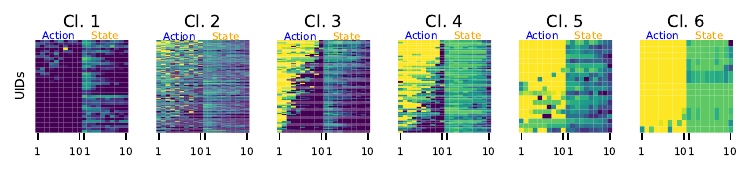}\\[1ex]
\fbox{\makebox[4.3cm][c]{\textbf{B.} Hierarchical Clustering}}\\
\includegraphics[width=\linewidth]{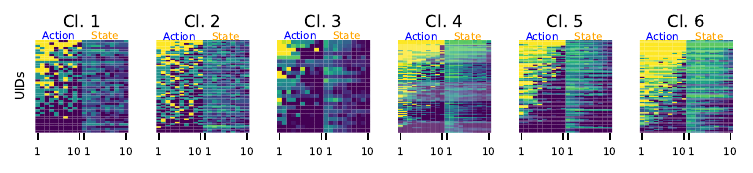}\\[1ex]
\fbox{\makebox[4.3cm][c]{\textbf{C.} C-Lasso}}\\
\includegraphics[width=\linewidth]{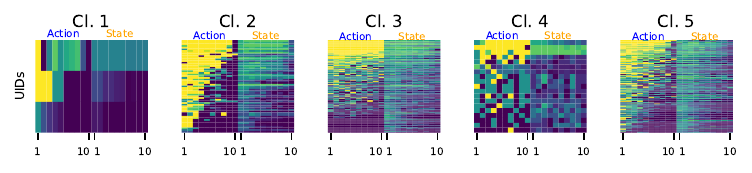}\\[1ex]
\fbox{\makebox[4.3cm][c]{\textbf{D.} Finite Mixture Model}}\\
\includegraphics[width=\linewidth]{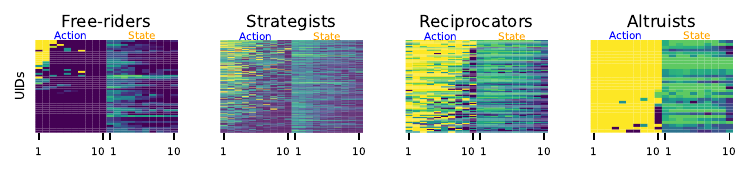}
\end{minipage}
}
\caption{{\bf Comparison of partitioning methods.} (A) Dynamic Time Warping (DTW) and Spectral Clustering, (B) Hierarchical Clustering \citep{Fallucchi2019}, (C) Classifier-Lasso \citep{su2016identifying}, and (D) Finite Mixture Models \citep{bardsley2007experimetrics}. We argue that DTW provides the sharpest separation of temporal patterns in our panel data.}
\label{fig:method_comp}
\end{figure}

Each method tested in Fig.~\ref{fig:method_comp} produces a distinct partition of the participant trajectories.

The finite mixture model is predefined to identify four behavioral types: freerider, altruist, reciprocator, and strategist. While the freerider and altruist clusters are clearly delineated, the reciprocator and strategist clusters are relatively large and noisy.

C-Lasso converged with five clusters rather than our proposed six. Its first cluster contains only three participants, whereas the others are comparatively large. Interestingly, Cluster 2 and Cluster 5 show similar temporal patterns. However, Cluster 3 mixes unconditional cooperators with consistent freeriders.

Hierarchical clustering produced six clusters but failed to recover the most prominent and interpretable types, namely the altruist and freerider groups.

By contrast, our DTW + Spectral Clustering approach generates clusters that are both temporally aligned and behaviorally coherent.

\clearpage
\section{Full Sample Plots}
\label{apx:full_sample}

\begin{figure}[H]
    \centering
    \includegraphics[width=0.99\linewidth]{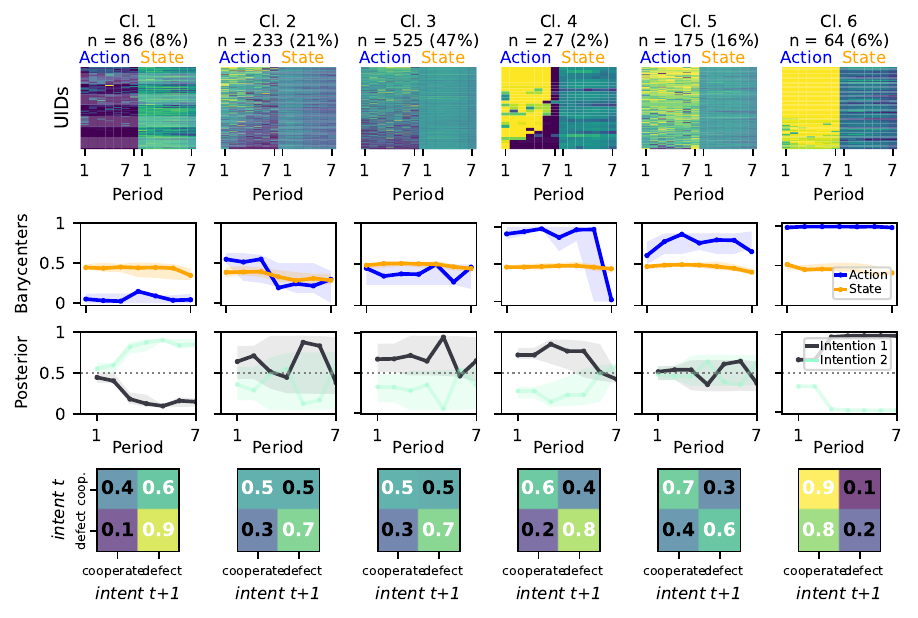}
    \caption{{\bf Clustering Analysis of Action-State Patterns and Intention Dynamics.} (1) Heatmaps of action and state trajectories per cluster, ordered by mean action.
(2) DTW barycenters of action-state patterns per cluster, with interquartile ranges shaded.
(3) DTW barycenter averages of posterior intentions per cluster, with interquartile ranges shaded.
(4) Average transition probabilities between cooperative and defecting intentions per cluster.}
    \label{fig:master_panel_7}
\end{figure}

Our dataset includes experimental games with varying lengths of play: 7, 10, 20, and 30 rounds. Each subset exhibits distinct structural and behavioral characteristics.
The 7-round subset is unique in that group sizes were exceptionally large—sometimes including up to 100 participants—which tends to stabilize average experiences across rounds. 
As a result, the state heatmaps appear markedly more uniform in this subset than in any of the others.
The 20-round subset differs in another important respect: it includes a punishment treatment. This feature likely explains the distinctive pattern observed for Cluster 3 in this condition (row 3), where contributions start low (dark purple) but subsequently rise to high levels (yellow)—a dynamic not seen in any other subset.

\begin{figure}[H]
    \centering
\includegraphics[width=0.99\linewidth]{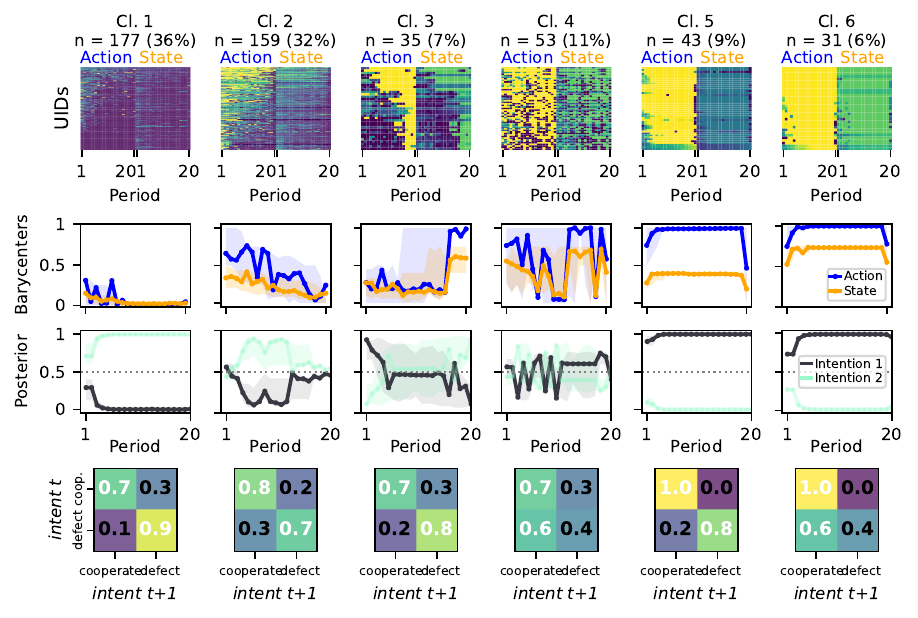} \\
\includegraphics[width=0.99\linewidth]{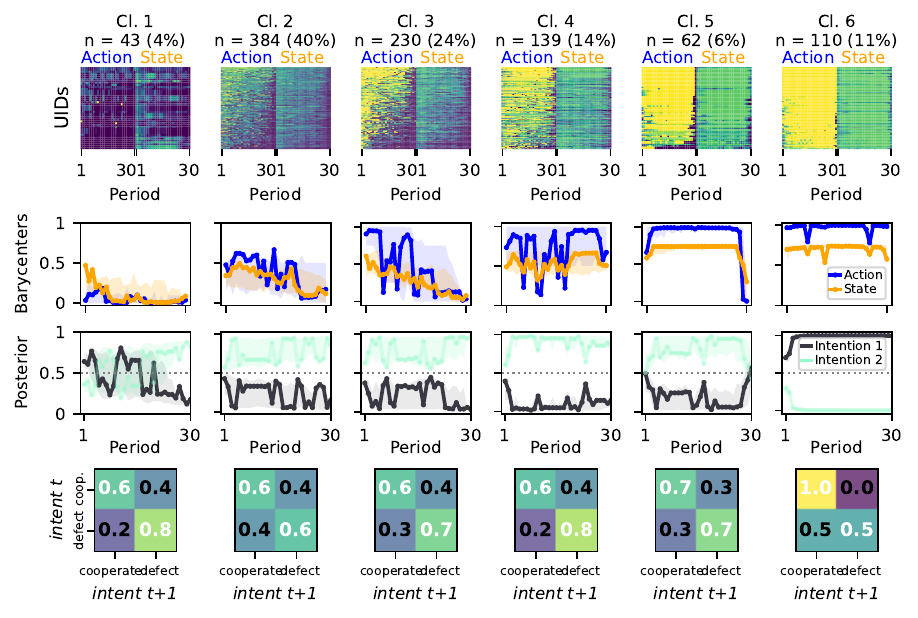}
    \caption{{\bf Clustering Analysis of Action-State Patterns and Intention Dynamics.} Additional panels for round lengths 20 and 30.}
    \label{fig:master_panel_20_30}
\end{figure}

\clearpage

\begin{figure}
    \centering
    \includegraphics[width=0.99\linewidth]{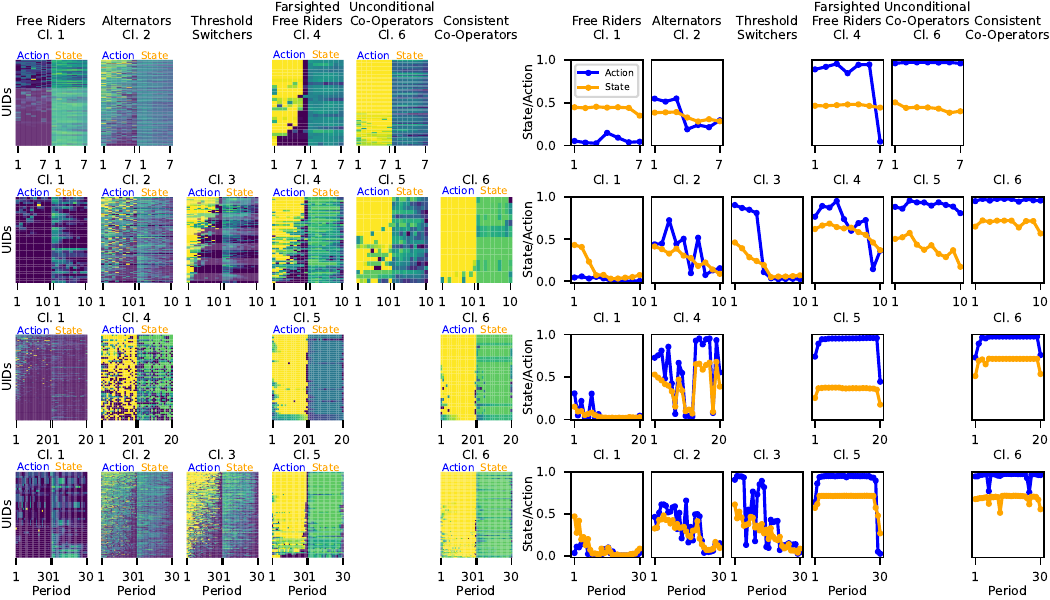}
    \caption{{\bf Behavioral Types Across Game Lengths.} The figure displays behavioral clustering results across four game lengths (7, 10, 20, and 30 periods). Each column represents a distinct behavioral type: Free Riders, Alternators, Threshold Switchers, Farsighted Free Riders, Unconditional Co-Operators, and Consistent Co-Operators. For each type, the left panel shows individual contribution patterns as heatmaps (with action and state contributions side-by-side), while the right panel displays the DBA-computed barycenter trajectories of actions and states over time. Rows correspond to different game lengths. Empty cells indicate that a particular type was not identified in that subset.}
    \label{fig:type-comparison}
\end{figure}

\clearpage

\begin{figure}
    \centering
\includegraphics[width=\linewidth]{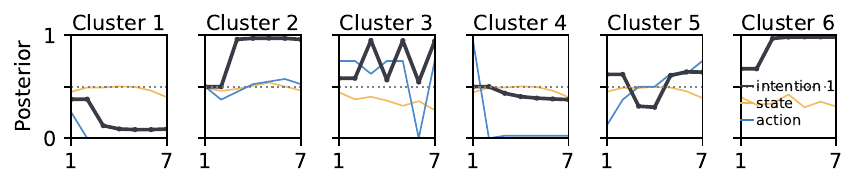}\\[1ex]
\includegraphics[width=\linewidth]{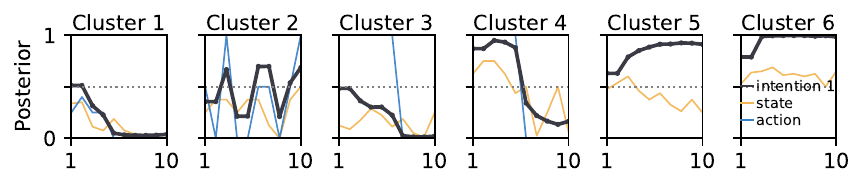}\\[1ex]
\includegraphics[width=\linewidth]{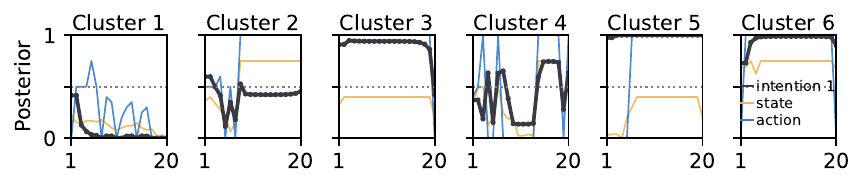}\\[1ex]
\includegraphics[width=\linewidth]{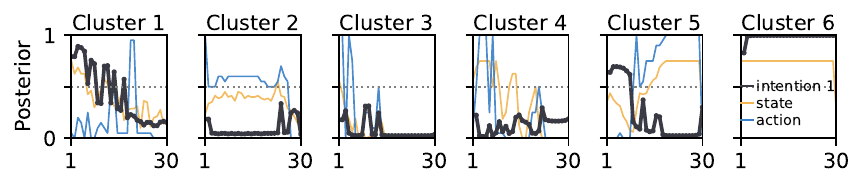}
    \caption{{\bf Sampled UIDs with Actions, States and Fitted Intention-Dynamics from Each Cluster.} Illustration of how observed actions and states inform latent intention adoption. For each cluster, we display randomly selected individuals and plot three aligned time series: their contributions (actions), their inferred states, and the posterior probability of adopting Intention 1. Because the model is based on two intentions, the probability of Intention 2 is simply one minus the plotted value.}
    \label{fig:samples}
\end{figure}

\clearpage
\section{Number of Latents and Log Likelihoods}

\begin{table}[H]
    \centering
\caption{Optimal Number of Latent Intentions: $K = 2$.}
\label{tab:likelihood_bic}

\begin{tabular}{c@{\hskip 0.8in}c@{\hskip 0.8in}c}
\toprule
\textbf{Transition ($K$)} & $\Delta$ Test LL & $\Delta$ BIC \\
\midrule
1 $\rightarrow$ 2 & 0.66 & 75.2 \\
2 $\rightarrow$ 3 & 0.31 & 88.7 \\
3 $\rightarrow$ 4 & 0.19 & 101.6 \\
4 $\rightarrow$ 5 & 0.19 & 114.4 \\
\bottomrule
\end{tabular}
\vspace{4pt}
\parbox{\linewidth}{\footnotesize\raggedright
\textit{Note:} Model comparison based on changes in test log-likelihood and Bayesian Information Criterion (BIC) when increasing the number of latent intentions ($K$). The transition from one to two intentions yields the largest improvement in test log-likelihood and the smallest rise in BIC, indicating that a two-intention model provides the best fit to the data.}
\end{table}

\section{Model Performance per Cluster and Subset}

%FILE LL_all.tex
% produced in IRL_C3_3

\begin{table}[H]
    \centering
\caption{Model Performance per Cluster and Subset.}
    \label{tab:cluster_loglikelihood}
\begin{tabular}{lllll}
\toprule
 & Subset 7 & Subset 10 & Subset 20 & Subset 30 \\
\midrule
Cluster 1 & -0.22 ± 0.04 & -0.24 ± 0.03 & -0.29 ± 0.02 & -0.38 ± 0.01 \\
Cluster 2 & -0.75 ± 0.02 & -0.72 ± 0.02 & -0.46 ± 0.06 & -0.81 ± 0.01 \\
Cluster 3 & -0.82 ± 0.02 & -0.59 ± 0.03 & -0.71 ± 0.02 & -0.74 ± 0.02 \\
Cluster 4 & -0.75 ± 0.11 & -0.88 ± 0.08 & -0.44 ± 0.04 & -0.53 ± 0.03 \\
Cluster 5 & -0.69 ± 0.02 & -0.67 ± 0.24 & -0.33 ± 0.06 & -0.61 ± 0.02 \\
Cluster 6 & -0.21 ± 0.06 & -0.33 ± 0.21 & -0.18 ± 0.05 & -0.12 ± 0.01 \\
\bottomrule
\end{tabular}
\vspace{4pt}
\parbox{\linewidth}{\footnotesize\raggedright
\textit{Note:} The model was fit across five cross-validation folds and ten repetitions. To assess model performance comprehensively, test log-likelihoods were first averaged across folds and repeats for each participant, and these participant-level means were then aggregated within each cluster. As the values represent log-likelihoods, less negative values indicate a better model fit to the human behavioral data.}
\end{table}

In IRL models, log-likelihood (LL) is used as a standard metric of model fit because it directly quantifies how probable the observed choices are under the fitted model—higher (less negative) values indicate that the model assigns greater probability to the actual behavior observed. However, it is important to note that LL not only measures model performance but also reflects the inherent noise in the behavioral data being modeled.
Table~\ref{tab:cluster_loglikelihood} shows differences in model fit across clusters. These differences in log-likelihoods capture underlying variation in behavioral noisiness. IRL models explain observed actions by assuming they arise from approximately optimal reward-driven behavior; therefore, the attainable log-likelihood depends on how consistent participants’ actions are with such a structure.

When participants behave deterministically or follow stable strategies—as in Cluster 1 ($-0.24 \pm 0.03$) and Cluster 6 ($-0.33 \pm 0.21$)—the model can closely reproduce their choices, yielding high (less negative) log-likelihoods. Conversely, when behavior is highly stochastic—such as in Cluster 2 ($-0.72 \pm 0.02$), where participants alternate unpredictably between extreme actions—the data contain more ``decision noise," and even the best-fitting model achieves lower likelihood. 
Hence, the fitted log-likelihood serves as an empirical indicator of behavioral noisiness: the more random the observed actions, the poorer the attainable fit \cite[for more details see][p.9]{zhu2025fitting}.

\clearpage
\section{Additional Analyses}

\begin{figure}[H]
    \centering
    \includegraphics[width=0.99\linewidth]{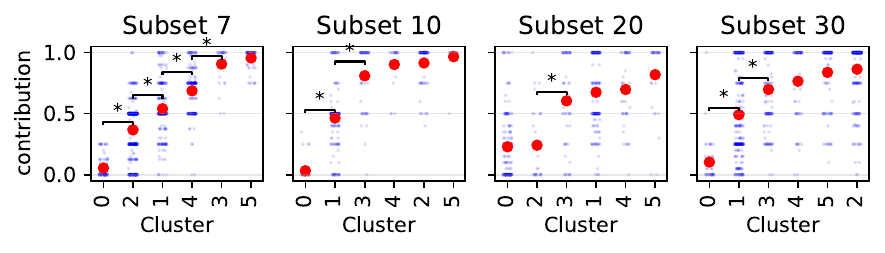}
    \caption{{\bf First Round Contributions by Cluster.} Each panel shows data from one subset size (7, 10, 20, or 30 participants). Within each subset, clusters are ordered along the x-axis according to their mean normalized contribution during the first round. Blue points represent individual participant contributions, and red circles indicate the mean contribution for each cluster. Adjacent clusters were compared using Welch's two-sample t-tests; black brackets mark pairs of neighboring clusters whose means differ significantly ($p <$ 0.05).}
    \label{fig:first_round_contr}
\end{figure}

First-round contributions are widely viewed as informative about underlying social preferences and early learning. Some work argues that these contributions matter chiefly in the initial round, when social motives are most salient, whereas later behavior is dominated by payoff-based learning. Motivated by this, we examined unconditional first-round contributions and the average contribution of group members by cluster (Fig.~\ref{fig:first_round_contr}). We tested for differences across clusters using Welch’s two-sample t-tests. Not all clusters differ significantly on either measure. Thus, initial contributions alone do not determine cluster membership; instead, clustering appears to reflect how individuals experience the interaction and adapt to those experiences, beyond their initial choices.

\end{appendices}
\end{document}